\documentclass[aps,pre,longbibliography,notitlepage,floatfix,nofootinbib,rmp,superscriptaddress]{revtex4-1}

\usepackage{bm,amsmath,amssymb,graphicx,stmaryrd,float,subcaption,upgreek,sidecap,MnSymbol,mathtools,lipsum}
\usepackage{epstopdf}
\usepackage{physics}
\usepackage[abs]{overpic}
\usepackage[justification=raggedright]{caption}
\usepackage{comment,footnote}
\usepackage{enumitem}
\usepackage{tikz}
\usepackage{hyperref}
\usepackage{wasysym}
\usepackage{relsize}
\usepackage{pifont}
\usepackage{xcolor}
\usepackage{tabularx}

\usepackage{etoolbox}
\patchcmd{\section}
{\centering}
{\raggedright}
{}
{}
\patchcmd{\subsection}
{\centering}
{\raggedright}
{}
{}

\newcommand{\graycircle}{{\color[rgb]{0.65,0.65,0.65}\text{\smaller[3]\CIRCLE}}}

\newcommand{\graytriangleL}{\color[rgb]{0.65,0.65,0.65} \text{\larger[1.0] $\!\!\!\blacktriangleleft$}}

\newcommand{\blackstar}{\color{black} \text{\smaller[4]$\bigstar$}}
\newcommand{\blackdiamond}{\color{black} \text{\smaller[2.0] $\!\!\blacklozenge$}}
\newcommand{\blackcircle}{{\color{black}\text{\smaller[10.0]\CIRCLE}}}
\newcommand{\blacktriangleL}{\text{\larger[1.0] $\!\!\blacktriangleleft\!$}}
\newcommand{\blacktriangleR}{\text{\larger[1.0] $\!\!\!\blacktriangleright$}}
\newcommand{\blacktriangleD}{ \text{\smaller[2.0] $\!\!\!\blacktriangledown$}}
\newcommand{\blasquare}{\color{black} \text{\smaller[1.0] $\!\!\blacksquare$}}
\newcommand{\blatriangle}{ \text{\smaller[2.0] $\!\!\!\blacktriangle$}}

\begin{document}
	\title{
Continuous modeling of creased annuli with tunable bistable and looping behaviors}
	\author{Tian Yu}
	\email{tiany@princeton.edu}
	\thanks{These two authors contributed equally }
	\affiliation{Department of Civil and Environmental Engineering, Princeton University, Princeton, NJ 08544}
	
	\author{Francesco Marmo}
	\email{f.marmo@unina.it}
	\thanks{These two authors contributed equally }
	\affiliation{Department of Structures for Engineering and Architecture, University of Naples Federico II, Naples, Italy}
	
	\author{Pasquale Cesarano}
	\affiliation{Department of Structures for Engineering and Architecture, University of Naples Federico II, Naples, Italy}
	
	\author{Sigrid Adriaenssens}
	\email{sadriaen@princeton.edu}
	\affiliation{Department of Civil and Environmental Engineering, Princeton University, Princeton, NJ 08544}
	\date{\today}

\begin{abstract}
Creases are purposely introduced to thin structures for designing deployable origami, artistic geometries, and functional structures with tunable nonlinear mechanics. Modeling the mechanics of creased structures is challenging because creases introduce geometric discontinuity and often have complex mechanical responses due to the local material damage. In this work, we propose a continuous description of the sharp geometry of creases and apply it to the study of \emph{creased annuli}, made by introducing radial creases to annular strips with the creases annealed to behave elastically. We find creased annuli have generic bistability and can be folded into various compact shapes, depending on the crease pattern and the overcurvature of the flat annulus. We use a regularized Dirac delta function (RDDF) to describe the geometry of a crease, with the finite spike of the RDDF capturing the localized curvature. Together with anisotropic rod theory, we solve the nonlinear mechanics of creased annuli, with its stability determined by the standard conjugate point test. We find excellent agreement between precision tabletop models, numerical predictions from our analytical framework, and modeling results from finite element simulations. We further show that by varying the rest curvature of the thin strip, dynamic switches between different states of creased annuli can be achieved, which could inspire the design of novel deployable and morphable structures. We believe our smooth description of discontinuous geometries will benefit to the mechanical modeling and design of a wide spectrum of engineering structures that embrace geometric and material discontinuities.
\end{abstract}

\maketitle

Abrupt changes are ubiquitous in nature and modern engineering, such as the jump of cross sections at a tree fork, impact forces, mantle discontinuities \cite{jeanloz1983phase}, the interfaces of shock waves \cite{fraley1986rayleigh}, and transient behaviors in signal processing \cite{bruce1996wavelet}, just to name a few.
These rapid changes correspond to a step-like jump or a localized spike, which can be described by a Heaviside function and a Dirac delta function, respectively.  
In mechanical engineering, material and geometric discontinuities have been introduced to elastic structures for various outcomes, such as optimal mechanical properties in stepped beams \cite{yang2004buckling} and remarkable stretchability in flexible electronics \cite{zhang2013buckling,zhang2014hierarchical}. 
Made by decorating thin sheets with certain crease patterns, origami structures bring great potential in achieving compact folding \cite{miura1985method,balkcom2008robotic,zirbel2013accommodating,wilson2013origami,wilson2013origami}, target geometries \cite{feng2020designs,schenk2013geometry,liu2019invariant,lang2012origami}, and tunable stiffness and stability in metamaterials  \cite{zhai2018origami,jules2022delicate,melancon2021multistable}.
However, it is challenging to study the mechanics of elastic structures with discontinuities (e.g. creases that correspond to $C^0$ continuity), which normally require cutting the structure and imposing matching conditions at the discontinuous interfaces \cite{dias2014non,filipov2017bar}.

In origami structures, creases have been modeled as discrete hinges \cite{lechenault2014mechanical,yu2021cutting,dias2014non,barbieri2019curvature,dharmadasa2020formation,yu2022avoiding}, smooth folds with $C^1$ continuity \cite{peraza2016kinematics,peraza2017design,walker2019flexural}, and as thinner \cite{andrade2019foldable,yan2016controlled} or narrower \cite{shi2017plasticity} structural elements, which either need a careful specification of matching conditions between the creases and the joining facets or require a detailed definition of the crease region. Recently, Jules et al. used the Heaviside feature of a hyperbolic tangent function to describe the local geometry of creases as $C^{\infty}$ continuity and studied the mechanics of creased \emph{elastica} \cite{jules19local}.

Here, we propose a $\Delta$ function that includes both a boxcar feature and a Dirac delta feature. We use the latter for a continuous description of the crease geometry and implement it with anisotropic rod theory to study the nonlinear mechanics of \emph{creased annuli}, which are found to have generic bistability and rich looping behaviors.
We find excellent agreement between precision tabletop models and numerical predictions from anisotropic rod theory, both showing that the bistability and looping behaviors can be tuned by varying the crease pattern and the overcurvature of the flat annulus. Overall, our work creates opportunities for folding strips into various shapes by introducing creases. The analytical framework could facilitate the mechanics design of thin structures embedded with discontinuities for desired mechanical and geometric properties.

\section*{$\Delta$ function}
 
In this work, we use the spike of a regularized Dirac delta function (RDDF) to model the localized curvature of a crease. We propose the following formulation consisting of two hyperbolic tangent functions,

\begin{equation} \label{eq:Deltafunc}
\begin{aligned}
\Delta_C ^{(l_b,l_e)}= \frac{1}{2(l_e - l_b)} \left[ \tanh \left(\frac{x-l _b }{C}\right)-\tanh\left(\frac{x-l _e }{C}\right)\right].
\end{aligned}
\end{equation}

\begin{figure}[tbhp]
	\centering
	\includegraphics[width=0.45\linewidth]{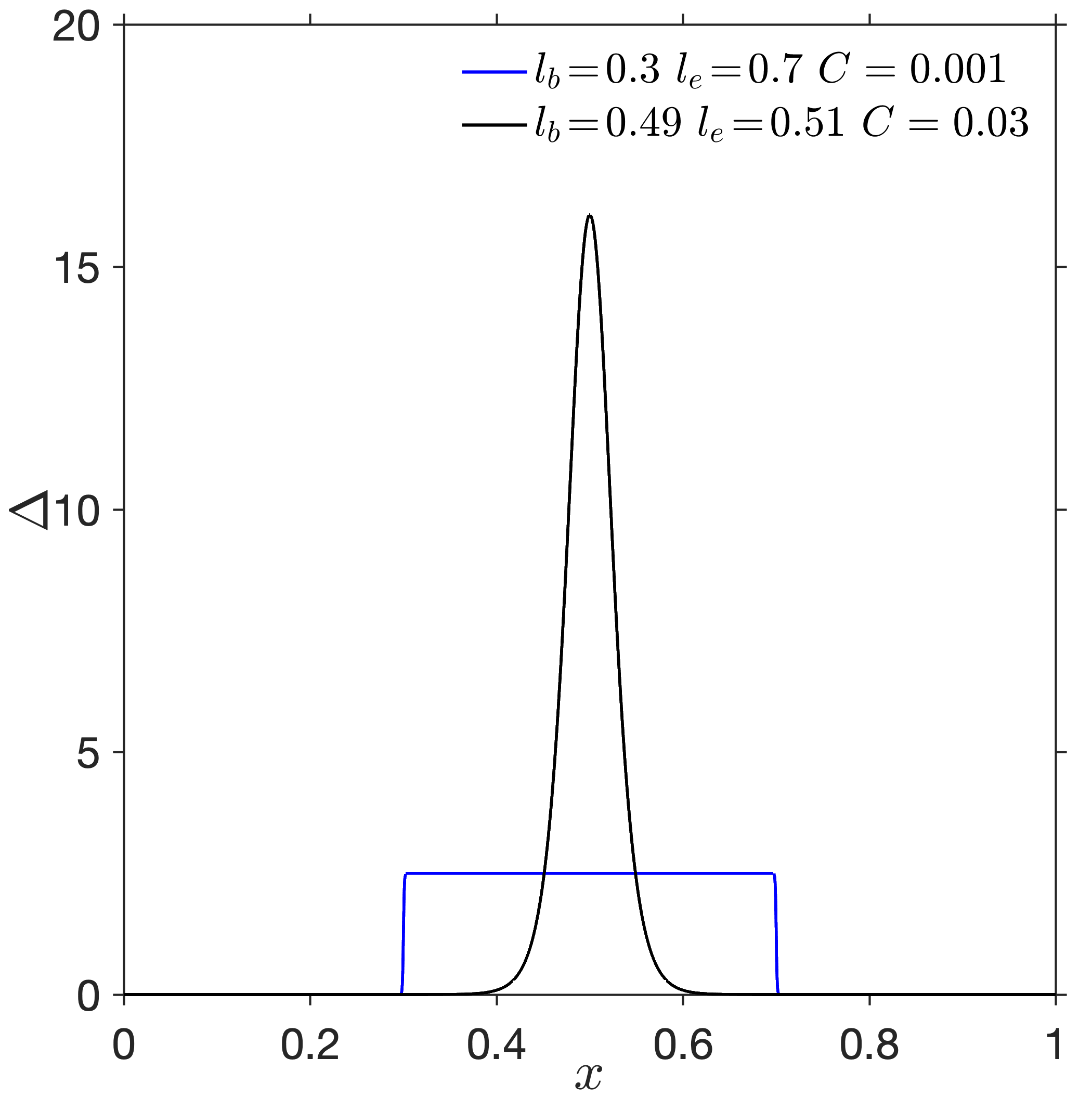}
	\caption{$\Delta$ in Equation \eqref{eq:Deltafunc} can mimic a boxcar profile (blue) and a regularized Dirac delta function (black).}
	\label{fig:DeltaFunction}
\end{figure}

First, we report several properties of $\Delta$ that may be used for the continuous description of discontinuities. When $C \! \ll \! (l_e -l_b)$, the two steps of hyperbolic tangents are separated and $\Delta$ appears to be a boxcar function, with its value being a constant $1/(l_e - l_b)$ in the range $x \in [l_b,l_e]$ and zero elsewhere (see the blue curve in Figure \ref{fig:DeltaFunction}). As $C \to 0$, $\Delta$ approaches a perfect boxcar function.
When $(l_e -l_b) < C \cup (l_e -l_b) \sim C$, the two steps of hyperbolic tangents collide and
$\Delta$ mimics a regularized Dirac delta function (RDDF), symmetric about and maximized at $x=(l_e + l_b)/2$ (see the black curve in Figure \ref{fig:DeltaFunction}). As $(l_e -l_b) \to 0$ and $C \to 0$, $\Delta$ approaches the Dirac delta function, with its value being infinite at $x=l_b=l_e$. In addition, it can be shown that $ \int _{-\infty} ^ {\infty} \Delta dx =1$ (SI Appendix, section 1). 
When we use $\Delta$ to model the localized curvature of creases, this plays an important role in obtaining target crease angles, which are determined by the integral of $\Delta$.

The above properties of $\Delta$ offer promise for describing various types of discontinuities. For example, using the boxcar feature of $\Delta$, we can describe any piecewise continuous function as a single continuous piece with $C^{\infty}$ (SI Appendix, section 2 reports an example characterizing the New York City skyline). In the rest of the paper, we use $\Delta$ as a RDDF to model the curvature of creases in elastic strips and investigate the nonlinear mechanics of \emph{creased annuli}.

\section*{Continuous description of creases/kinks in anisotropic rods}

\begin{figure}[h!]
	\centering
	\includegraphics[width=0.75\linewidth]{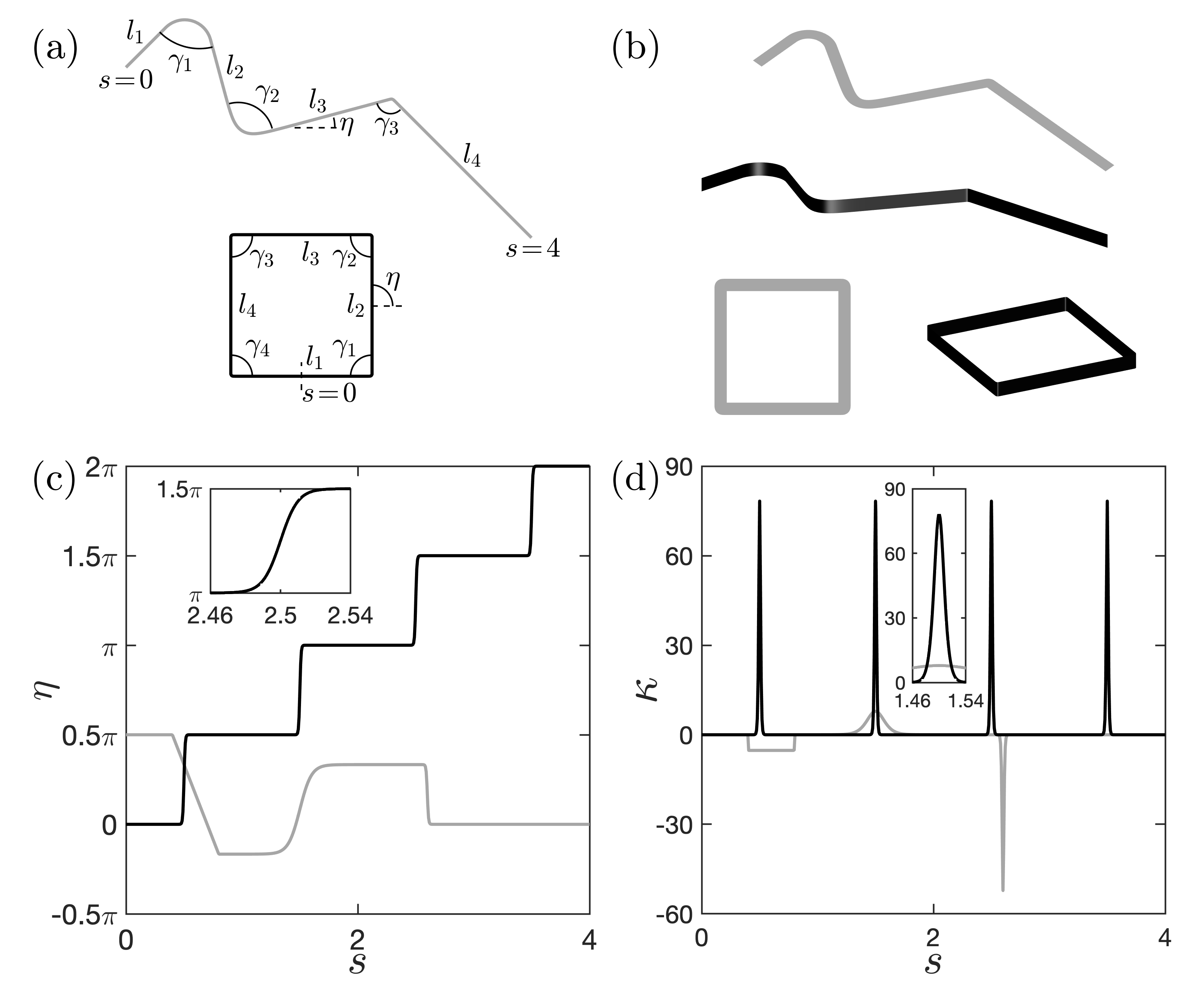}
	\caption{Strips with creases/kinks with the rest curvature described by Equation \eqref{eq:accordionsquare}. (a) Rods with multiple creases. The accordion has $n_c=3$, $(\gamma_1,C_1,l_{b1},l_{e1})=(\pi / 3, 0.001,0.4,0.8)$
		, $(\gamma_2,C_2,l_{b2},l_{e2})=(\pi / 2, 0.1,1.499,1.501)$, and $(\gamma_3,C_3,l_{b3},l_{e3})=(2\pi / 3, 0.01,2.599,2.601)$. The square has $n=4$ and $(\gamma_i,C_i,l_{bi},l_{ei})=(\pi / 2, 0.01,i-0.501,i-0.499)$.  (b) The planar rods in (a) are used as centerlines to construct strips with kinks (grey) and creases (black). (c-d) report the tangent angle $\eta$ and the curvature $\kappa$ of the rods in (a), respectively.}
	\label{fig:DeltaIntro}
\end{figure}

Perfect creases/kinks have a vanishing extension and correspond to $C^0$ continuity, i.e. the local tangent is discontinuous that results in a blow up of the local curvature. Due to the material thickness, creases in thin elastic structures always have a small extension, which results in the localization of the curvature. 
Here, we combine a series of $\Delta$ functions to describe the rest curvature of anisotropic rods with multiple creases/kinks,

\begin{equation} \label{eq:accordionsquare}
\begin{aligned}
\kappa&=\sum_{i=1}^{n_c} \text{sgn}_i \frac{\pi - \gamma_i}{2 (l_{ei} -l_{bi} ) } \left[ \tanh \left(\frac{s-l_{bi} }{C_i} \right)-\tanh\left(\frac{s-l_{ei} } {C_i} \right)\right] \\
&=\sum_{i=1}^{n_c} (-1)^{i} (\pi - \gamma_i)   \Delta_{C_i} ^{(l_{bi},l_{ei})} \,, \\
\end{aligned}
\end{equation}

where $n_c$, $\gamma_i$, $s$ ($\in [0,l]$), and $l$ represent the number of creases, the $i_{\text{th}}$ crease angle (see Figure \ref{fig:DeltaIntro}a for the definition of crease angle), the arc length of the rod, and its total length, respectively. The prefactor sgn$_{i}$ could be $\pm 1$ and is used to prescribe the bending direction of the crease, e.g. the grey accordion in Figures \ref{fig:DeltaIntro}a. 
$C_i$ and $(l_{ei} - l_{bi})$ determine the local crease profile centered at $s \!=\! (l_{ei} + l_{bi}) /2$. With $(l_{ei} - l_{bi}) \! \to \! 0$, our approach degenerates to Jules et al.'s method that uses a hyperbolic tangent to describe the local tangent angle of creases \cite{jules19local} (SI Appendix, section 2). With $C_i \! \ll \! (l_{ei} -l_{bi})$, \eqref{eq:accordionsquare} generates uniform bends in the regions $s \in [l_{bi}, l_{ei}]$ with constant curvatures. 

We define ($4 C_i+ l_{ei} - l_{bi}$) as the \emph{nomimal crease length}, which results in a crease region $s \in [l_{bi} - 2 C_i, l_{ei}+  2 C_i]$ and in which most of the crease angle $\gamma_i$ is achieved. 
Decreasing $C_i$ and $(l_{ei} - l_{bi})$ will sharpen the creases and make them independent of other creases. With the increase of ($4 C_i+ l_{ei} - l_{bi}$), adjacent creases will overlap and generate nonflat facets. As long as $(4C_i + l_{ei}- l_{bi})$ is small compared with the length of the adjacent facets, the multiple creases represented by Equation \eqref{eq:accordionsquare} do not affect each other and behave independently. A comprehensive discussion about the influences of $C_i$ and $(l_{ei} - l_{bi})$ on the crease profile and the errors of crease angles are included in SI Appendix, section 1.

The geometry of the rod is obtained by first integrating $\kappa$ to get the local tangent angle $\eta$ and then integrating the kinematic equations to obtain the rod profile. Figure \ref{fig:DeltaIntro}a displays two examples with the rod normalized to the same length $4$: a grey accordion with three unevenly distributed creases of different profiles and a black square with four evenly distributed creases of the same profile. $\eta$ measures the local tangent angle. The grey accordion contains a uniform crease with an angle $\gamma_1$ and two nonuniform creases of different profiles. 
The two examples in Figure \ref{fig:DeltaIntro}a are used as centerlines to construct strips with kinks (the width of the strip is coplanar to the centerline) and creases (the width of the strip is perpendicular to the centerline), as shown in Figure \ref{fig:DeltaIntro}b.

Figures \ref{fig:DeltaIntro}(c-d) present the distribution of the tangent angle $\eta$ and the rest curvature $\kappa$ of the two configurations in Figure \ref{fig:DeltaIntro}a, respectively. At a sharp crease, $\eta$ approaches a jump resulting in a spike in $\kappa$. Increasing the value of $(4C_i+l_{ei}-l_{bi})$ leads to blunter creases with the jump of $\eta$ and the spike of $\kappa$ being smoothed in the horizontal direction. In the remainder of this article, we apply the above framework to study the bistable and looping behaviors of \emph{creased annuli}.

\section*{Geometry of creased annuli and fabrication of tabletop models}

\begin{figure*}[h!]
	\centering
	\includegraphics[width=0.9\linewidth]{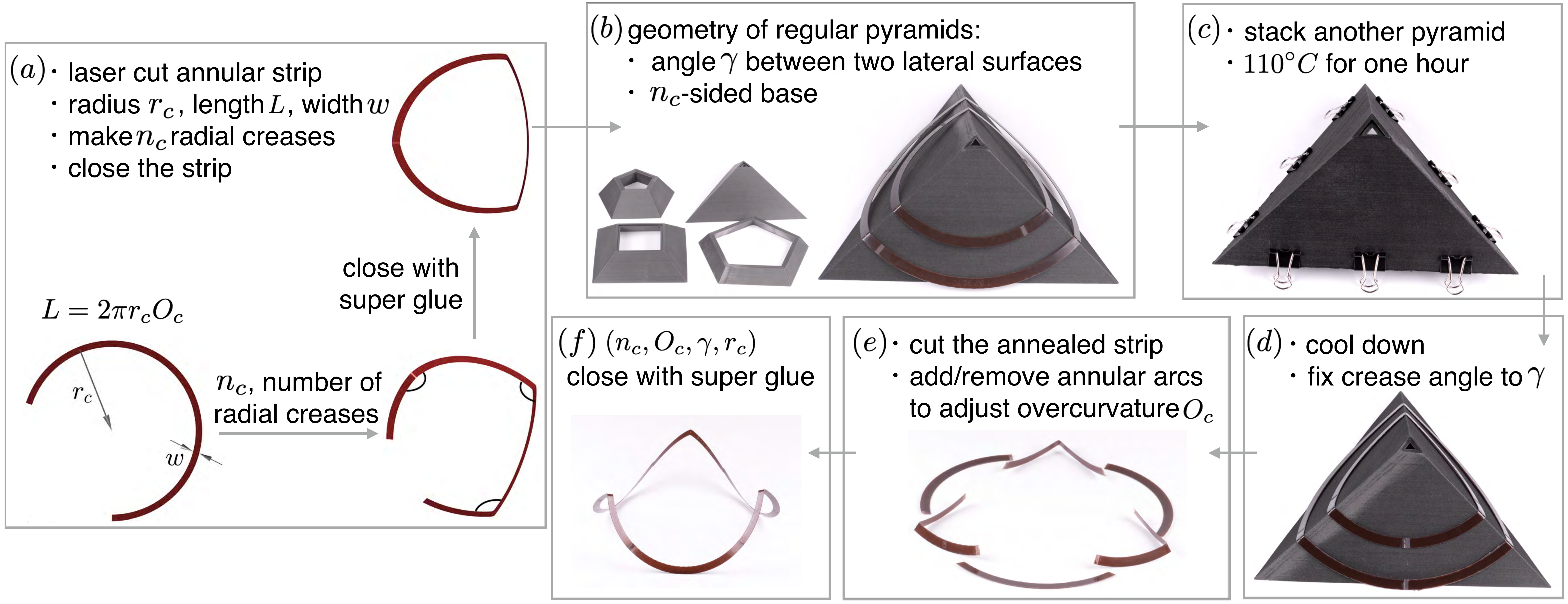}
	\caption{Geometry and fabrication of creased annuli. (a) A creased annulus is determined by four geometric parameters $(n_c, O_c, \gamma, r_c)$. (b) 3D printed pyramids that match with the geometry of creased annuli. (c-d) Annealing. (e) Adjust overcurvature. (f) Creased annuli with target geometric parameters. }\label{fig:Crease_Expts_Meth}
\end{figure*}

Creased annuli are made by first introducing evenly distributed radial creases of angle $\gamma$ to annular strips (with length $L$ and radius of the centerline $r_c$) and then forcing the two ends to close (Figure \ref{fig:Crease_Expts_Meth}a). We define \emph{overcurvature} as $O_c= L/(2 \pi r_c)$, which measures the number of loops the annuli cover in its flat rest state. The geometric parameters of the annular strips include number of creases $n_c$, overcurvature $O_c$, crease angle $\gamma$, and the radius $r_c$. 

The fabrication process is summarized in Figure \ref{fig:Crease_Expts_Meth}. We first laser cut annular strips with thickness $t \!=\! 0.254$ mm, width $w \!=\! 5.08$ mm, and length $L$ from polyester shim stock (Artus Corp., Englewood, NJ), make $n_c$ radial creases at the desired locations that have been slightly scored by the laser, and close the strip with super glue (Figs. \ref{fig:Crease_Expts_Meth}a). 
Then, we place the closed strip between two 3D printed pyramids with specific geometries \cite{yu2022bistability}, such that the thin strip closely matches the surfaces and ridges of the pyramids (Figs. \ref{fig:Crease_Expts_Meth}(b-c)). Next, to anneal the strip, we put the pair of stacked pyramids into an oven at $110^{\circ}C$ for one hour and cool the sample down at room temperature ($\approx 21 ^{\circ}C$), which fixes the crease angle to be the ridge angle of the pyramids (i.e. the dihedral angle between two lateral surfaces), eliminates the residual stress inside the crease, and leads to purely elastic creases \cite{klett2018paleo,sargent2019heat,jules19local} (Figs. \ref{fig:Crease_Expts_Meth}(c-d)).  
After annealing, we cut the strip and insert (remove) additional (redundant) flat annular strips to adjust the overcurvature (Figure \ref{fig:Crease_Expts_Meth}e). Finally, the structure is closed with super glue to obtain creased annuli with target geometry $(n_c, O_c, \gamma, r_c)$ (Figs. \ref{fig:Crease_Expts_Meth}f). Details of the geometry of the pyramids and the fabrication process are documented in SI Appendix, section 3.  
We find creased annuli have generic bistability with small overcurvature (Supplementary Video 1) and can be folded into various shapes by increasing the overcurvature (Supplementary Video 2).

\section*{Implementation with anisotropic rod theory}

Anisotropic rod theory is normally used to model slender rods or strips with a mild anisotropy of the cross section \cite{mahadevan1993shape} (i.e. $L>>w \sim t$). Several recent studies have demonstrated its accuracy in predicting the nonlinear mechanics of strips with $w/t$ up to $ O(10)$ \cite{yu2019bifurcations,riccobelli2021rods,moulton2018stable}. Throughout this study, we fix $w/t$ to 20 and use the anisotropic rod model to study creased annuli. The force and moment balances of an inextensible and unshearable rod can be summarized as  

\begin{equation}\label{eq:F&Mbalance}
\begin{aligned}
\bm{N}' &=\bm{0} \,, \bm{M}' +\bm{d_3} \times \bm{N} &=\bm{0} \,, \\
\end{aligned}
\end{equation}

where a prime denotes an $s$ derivative ($s$ ($\in [0,l]$)), and $\bm{N}$ and $\bm{M}$ represent contact forces and moments, respectively. $\bm{d}_3$ corresponds to the unit tangent vector on the centerline of the strip.
We assume linear constitutive relations $M_1=EI_1 (\kappa_1 - \kappa_{10})$, $M_2=EI_2 (\kappa_2 - \kappa_{20})$, and $M_3= GJ \tau$, where the contact moment $M$ has been resolved on a local material frame (SI Appendix, section 4). $E$ and $G$ are the Young's modulus and shear modulus of the constituent isotropic material, respectively; $EI_1$ and $EI_2$, and $GJ$ correspond to the two bending rigidities and the torsional rigidity, respectively. $\kappa_1$, $\kappa_2$ and $\tau$ are the curvatures and twist in the deformed configuration; $\kappa_{10}$ and $\kappa_{20}$ represent the rest curvatures, which correspond to the geodesic curvature of the annulus and the localized curvature of the creases, respectively. Here, we have $\kappa_{10}=2 \pi O_c /L$ and $\kappa_{20}=\sum_{i=1}^{n_c} (\pi - \gamma_i)   \Delta_{C_i} ^{(l_{bi},l_{ei})} $. In addition, the length of the strip $l$ in numerics is normalized to $n_c$.

To model the sharp creases of creased annuli, we fix $(l_{ei} - l_{bi})$ to a small value $2 \times 10^{-4}C_i$, which results in a nominal crease length $4C_i$ (notice that $(l_{ei} - l_{bi}) << C_i$). In tabletop models, the crease length mainly depends on the thickness $t$ of the material and has been shown to be in the order $O(t) \sim O(10t)$ \cite{benusiglio2012anatomy,jules19local}.
In order to make the sharpness of the crease in the numerical modeling be realistic compared with experimental models, we set $10t/(2 \pi r_c O_c )=4C_i /l$, where we have estimated the crease length to be $10t$ and set the ratio between the crease length and the total strip length to be the same for experiments and numerics. For all the tabletop models, the corresponding $C_i$ is found to be in $[8.42 \times 10^{-4},1.03 \times 10^{-2}]$. 

Notice that we have only set the length of the crease in numerics to be approximately the same as the crease length in experiments; our continuous description of the crease geometry through the specification of $\kappa_{20}$ does not necessarily match with the local crease profile in experiments, which could depend on the material and the creasing method. In SI Appendix, section 5, we show that with sharp creases, which are typical for creased thin sheets, the effects of the local crease geometry on the numerical results are negligible.    
After nondimensionalization, the only material parameter is the Poisson's ratio $\nu$, which we set to 0.33 for the current study. Details of the anisotropic rod model, its implementation with numerical continuation package AUTO 07P \cite{doedel2007auto} for solving static equilibria, and the stability test of the equilibria are discussed in SI Appendix, section 4. Numerical continuation is powerful for conducting parametric studies and can trace the solutions as a bifurcation parameter varies. AUTO 07P is able to detect various kinds of bifurcations and folds and further compute the bifurcated branches \cite{doedel2007auto}.

\section*{Creased annuli with tunable bistability and looping behaviors }

\begin{figure*}[h!]
	\centering
	\includegraphics[width=0.95\linewidth]{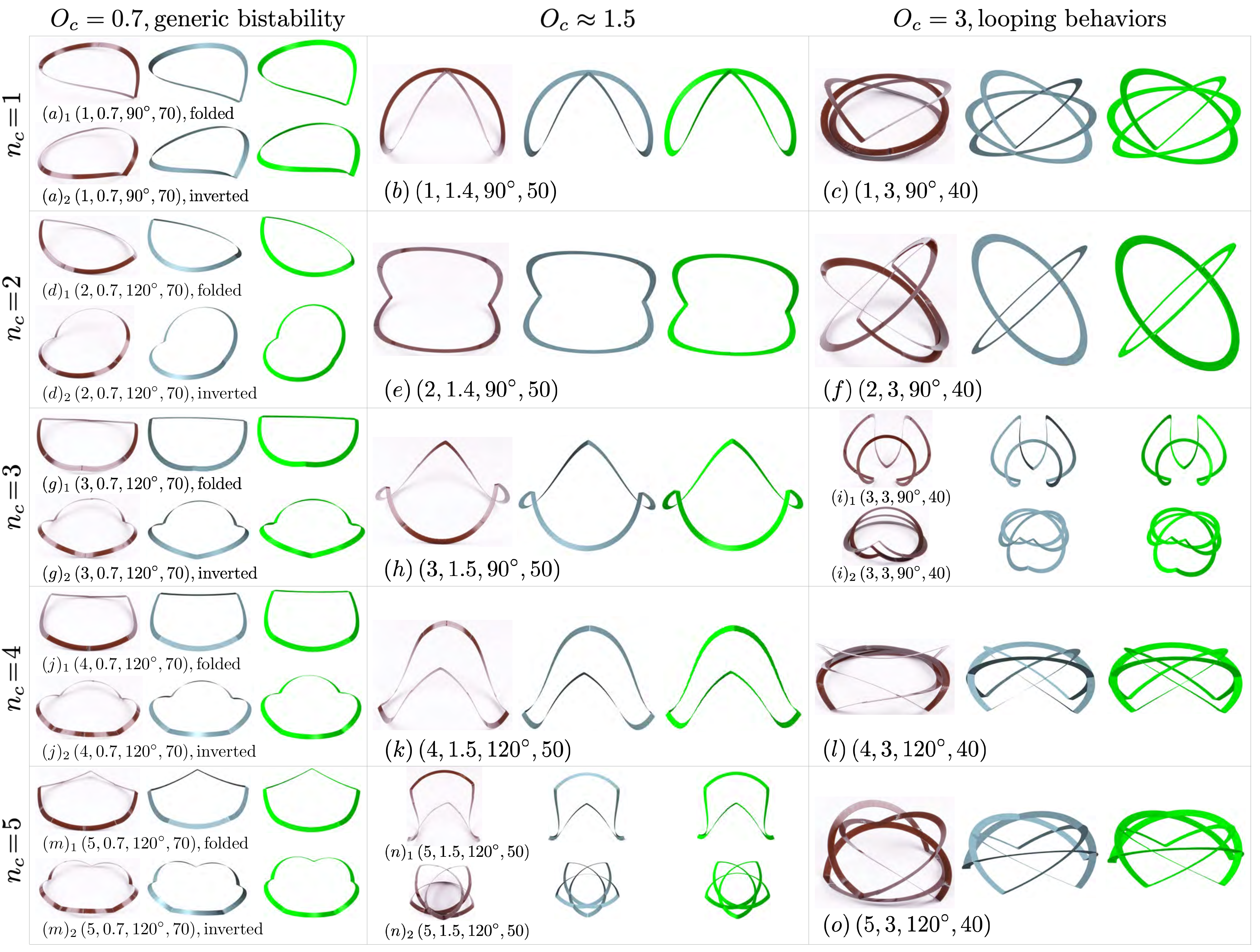}
	\caption{Comparison of experimental models of creased annuli (brown), stable states from numerical predictions of anisotropic rod (blue), and equilibria obtained from finite element simulations (green), with different geometric parameters $(n_c, O_c, \gamma, r_c)$. $()_{1,2}$ correspond to a bistable pair. Generic contact is observed in looped states and is not included in numerical modelings.}\label{fig:CreaseAnnuliConfig}
\end{figure*}

Comparisons between the numerical results (blue renderings) and experimental models (brown) are summarized in Figure \ref{fig:CreaseAnnuliConfig} with different geometric parameters $(n_c, O_c, \gamma, r_c)$. We find excellent agreement, except for the looped configurations where contact exists in physical models and is not included in numerical predictions.
The blue renderings are constructed from the anisotropic rod model and have the same slenderness (i.e. $w/L$) with the corresponding experimental models. We additionally conduct finite element (FE) simulations using the commercial software ABAQUS, with the modeling results presented in Fig. \ref{fig:CreaseAnnuliConfig} as green shapes. Creases and overcurvature of creased annuli are generated by applying temperature gradients along both the thickness and width of the cross section (Materials and Methods). All the FE results agree well with our numerical predictions from anisotropic rod theory. This confirms the accuracy of our framework and further implies that self contact could be the main cause of the differences between the experimental models and the numerical results for the looped configurations. Parametric studies with ABAQUS show that material properties (i.e. Young's Modulus and Poisson's ratio) have negligible effects on the static equilibria of creased annuli. These findings agree with the conclusions drawn from our theoretical framework (SI Appendix, section 4).

We find that creased annuli have generic bistability with small overcurvatures and tunable looping behaviors with large overcurvatures. In the first row of Figure \ref{fig:CreaseAnnuliConfig} ($O_c \!=\! 0.7$), $()_1$ and $()_2$ represent a bistable pair and are referred to as \emph{folded} and \emph{inverted} state respectively. For example, $(a)_1$ and $(a)_2$ are a bistable pair that can be manually deformed to one or the other (Supplementary Video I). 
In the third row of Figure \ref{fig:CreaseAnnuliConfig} with a large overcurvature $O_c \!=\! 3$, creased annuli fold into multiply covered loops, e.g. in (c), (f), $(i)_2$, (l), and (o). Notice that with three creases, the looped configuration in $(i)_2$ resembles a triply-covered version of $(a)_1$ and could further be deployed to a stable flower-like shape in $(i)_1$. 
With an intermediate overcurvature around 1.5 (the second row in Figure \ref{fig:CreaseAnnuliConfig}), creased annuli with one, two, there and four creases are monostable, corresponding to $(b)$, $(e)$, $(h)$ and $(k)$, respectively; the creased annulus with five creases is bistable, which can be folded into a star configuration $(n)_2$. It is known that annular strips without creases will fold into multiply covered loops at an overcurvature of odd integers (i.e. $O_c=3,5,7...$) \cite{manning2001stability,audoly2015buckling}. Our results show that by introducing radial creases, the out-of-plane mechanical behaviors of annular strips could be significantly enriched, creating various folding patterns and stable configurations that could be tuned by the number of creases and overcurvatures.

\begin{figure*}[h!]
	\centering
	\includegraphics[width=0.95\linewidth]{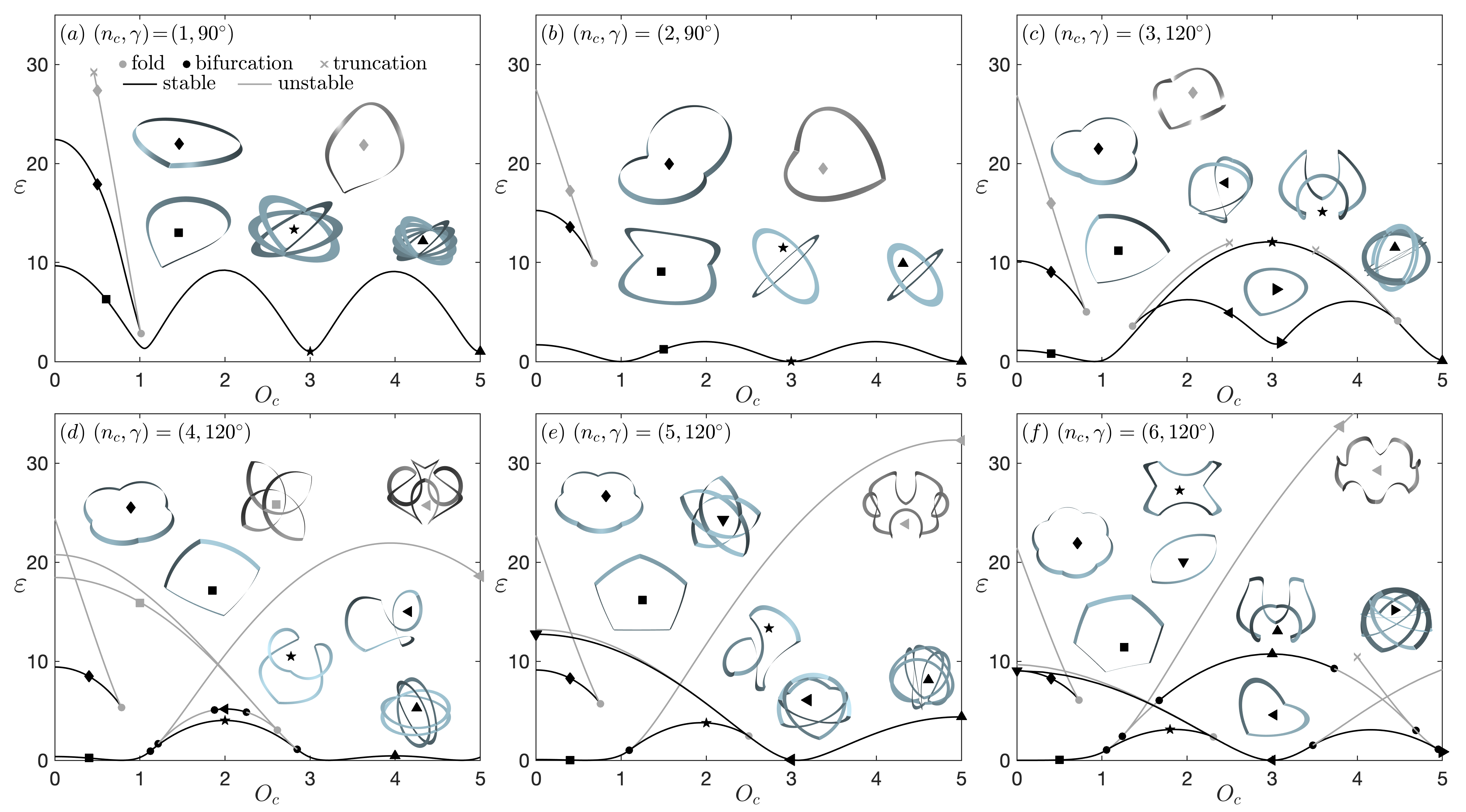}
	\caption{Bifurcation diagrams of creased annuli with fixed crease angle $\gamma$ and different number of creases $n_c$. Black and grey curves correspond to stable and unstable solutions, respectively. }\label{fig:AnnularBifurcation}
\end{figure*}

We further investigate how overcurvature affects the nonlinear mechanics of creased annuli through numerical continuation. Figure \ref{fig:AnnularBifurcation} reports the bifurcation diagram of creased annuli with different number of creases in the $O_c - \varepsilon$ plane, with $\varepsilon \!=\! 0.5  \int _0 ^{l} [a (\kappa_1-\kappa_{10}) ^2 +b (\kappa_2-\kappa_{20}) ^2 +\tau^2] \, d s $ corresponding to the normalized elastic energy and $a$ and $b$ being the two normalized bending stiffnesses (SI Appendix, section 4). Here we have fixed $(C_i,l_{ei} - l_{bi})$ to $(0.002,2 \times 10^{-4} C_i)$. Black and grey solid circles represent bifurcation and fold points, respectively. Some of the unstable branches are partially reported here and truncated by a grey cross.
Renderings represent solutions marked on the curves. Grey and black curves correspond to unstable and stable solutions, respectively. The nonlinear stability of equilibria is obtained by conducting the conventional conjugate point test for a single rod (SI Appendix, section 4).

The inverted branch (marked by $\blackdiamond$) loses stability through a fold point with $O_c <1$ for $n_c=2,3,4$ and 5 and $O_c \! \approx \! 1$ for $n_c=1$. On the other hand, with the increase of the overcurvature, the folded shape $\blasquare$ could directly lead to looped states for $n_c=1,2$ and 3 following the $\blasquare \!\! \to \!\! \blackstar \!\! \to \!\! \blatriangle$ branch, or lose stability through bifurcations for $n_c=4, 5$ and 6 following the $\blasquare \!\! \to \!\! \blackcircle \!\! \to \!\!  \blackstar$ branch, with the bifurcated branch $\blackstar$ connected to looped configuration for $n_c \!=\!4$ or losing stability through a fold point for $n_c \!=\! 5$ and 6. In the latter, looped states are independent branches that are disconnected from the folded state $\blasquare$. For example, the $\blacktriangleD \!\! \to \!\! \blacktriangleL \!\! \to \!\! \blatriangle$ branch for $n_c \!=\! 5$ and the $\blacktriangleD \!\! \to \!\! \blacktriangleL \!\! \to \!\! \blackcircle$ branch for $n_c \!=\! 6$ represent multiply-covered branches. Stable states could also exist in a stability island, for example, the $\blacktriangleL \! \to \! \blacktriangleR$ branch for $n_c \!=\! 3$ (bounded by two fold points), the $\blacktriangleL$ branch for $n_c \!=\! 4$ (bounded by two bifurcations), the $\blatriangle$ branch for $n_c \!=\! 6$ (bounded by two bifurcations), and the $\blacktriangleR$ branch for $n_c \!=\! 6$ that gains stability through a bifurcation.
Generally speaking, increasing the overcurvature $O_c $ will fold creased annuli into different multiply covered shapes, depending on the number of creases. The looped states normally contain less elastic energy $\varepsilon$. In addition for $n_c \geq 3$, increasing $O_c$ deforms the folded branch ($\blasquare$) into a flower-like shape, which is stable for $n_c=3$ ($\blackstar$) and unstable for $n_c=4,5$ and 6 ($\graytriangleL$).

\pagebreak
\section*{Switches between different states} 

\begin{figure*}[h!]
	\centering
	\includegraphics[width=0.85\linewidth]{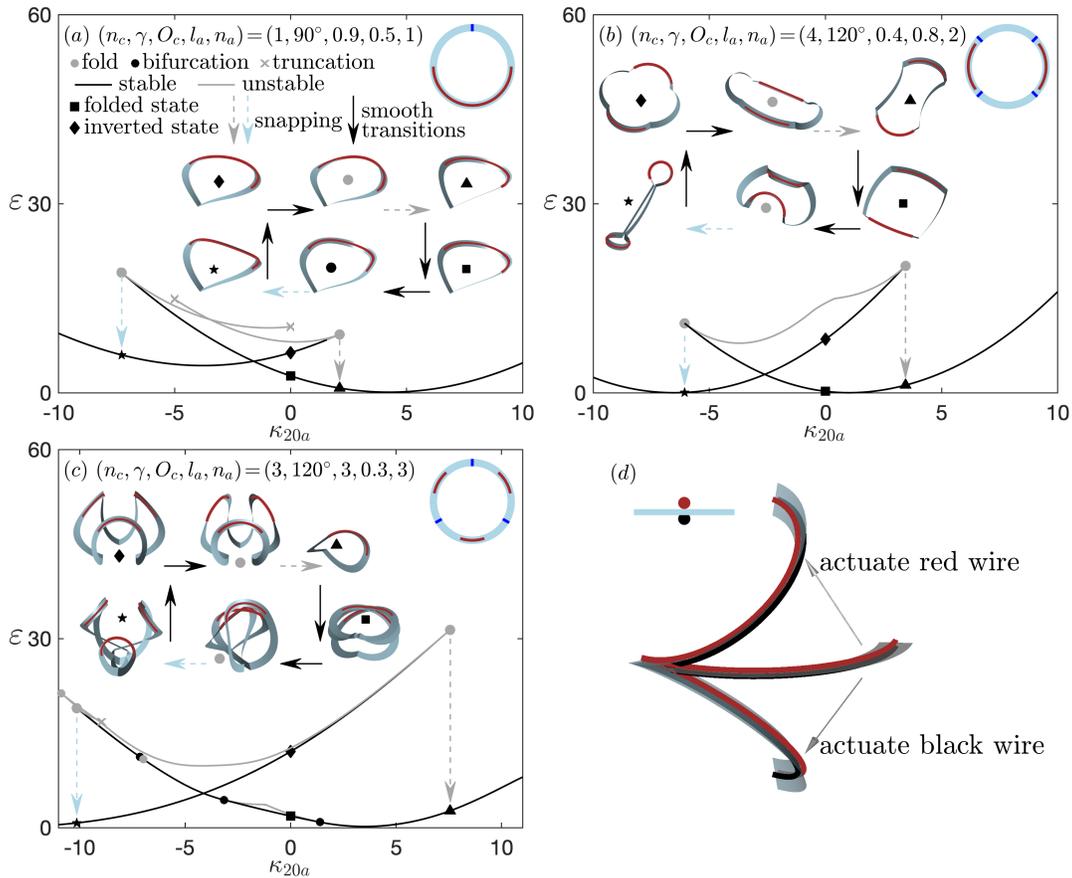}
	\caption{State switches between different states of creased annuli are achieved by introducing an actuation curvature $\sum_{i=1}^{n_a} \kappa_{20a} \Delta_{C}^{(l_{bi},l_{ei})}$ in the direction of the minimum bending stiffness of thin strips. (a-c): Bifurcation diagrams illustrating state transitions. The blue and red lines in the layout at the top right of each figure represents creases and actuated segments, respectively. (d) Schematics of SMA actuation method. Actuating smart memory alloy wires attached to strip's upper and bottom surfaces will bend the strip, which appears to introduce the actuation curvature. The inset shows the cross section of the strip.}
	\label{fig:StateSwitch}
\end{figure*}

Applications of the bistable and looping behaviors in creased annuli may require remote switches between different states. For a thin strip, it is much easier to bend it about the width direction than about the surface normal. Here we show that adding an actuation curvature $\sum_{i=1}^{n_a} \kappa_{20a} \Delta_{C}^{(l_{bi},l_{ei})}$ in the direction of the minimum bending stiffness to $\kappa_{20}$ could trigger dynamic snappings between different states. We adopt the boxcar feature of the $\Delta$ function by setting $C << (l_{ei}-l_{bi} )$; $n_a$ represents the number of actuated segments and $l_{bi}$ and $l_{ei}$ represent the beginning and the end of the $i_{\text{th}}$ actuated segment, respectively. This allows us to freely vary the number, length, and location of the actuated segments. \\
Figs. \ref{fig:StateSwitch}(a-c) display several bifurcation diagrams under the application of actuation curvatures. Each figure contains a layout at the top right showing creases (blue lines) and actuated segments (red lines), which have the same length $l_a$ ($=l_{bi}-l_{ei}$) and are placed symmetrically between adjacent creases. The vertical axis $\varepsilon$ reports the normalized elastic energy in the structure, namely $ 0.5  \int _0 ^{l} [a (\kappa_1-\kappa_{10}) ^2 +b (\kappa_2-\kappa_{20}-\sum_{i=1}^{n_a} \kappa_{20a} \Delta_{C}^{(l_{bi},l_{ei})} ) ^2 +\tau^2] \, d s $. 
The solution curves are similar in all three examples: the structure loses stability through a fold point with the increase of $|\kappa_{20a}|$, followed by a dynamic jump to a stable state with a lower energy level. Fig. 6a shows the transition between the inverted and the folded state of a creased annulus with a single crease. If we start with the inverted state $\blackdiamond$ and apply a positive $\kappa_{20a}$ to half of the strip, the structure loses stability at a fold point and jumps to the folded branch $\blatriangle$. With the deactivation of $\kappa_{20a}$, the structure follows the folded branch to reach the folded state $\blasquare$. On the other hand, if we start with the folded state $\blasquare$ and apply a negative $\kappa_{20a}$, the structure loses stability through another fold point and snaps back the inverted branch $\blackstar$. With the deactivation of $\kappa_{20a}$, the structure follows the inverted branch to reach the inverted state $\blackdiamond$.
This actuation sequence can be applied repeatedly to produce cyclic state switches between the inverted and the folded state.
Fig. \ref{fig:StateSwitch}b demonstrates similar transitions for a creased annulus with four creases by partially actuating two segments.
Fig. \ref{fig:StateSwitch}c reports the looping and deployment processes of a creased annulus with three creases, achieved by actuating three short segments. \\
Notice that there could be many actuation options for achieving transitions between different states of creased annuli. We provide a robust framework that is convenient for designing and further optimizing the actuation scheme because the boxcar feature of the $\Delta$ function enables us to freely vary the number, length, and location of the actuated segments. In engineering applications, the actuation curvature can be realized by activating smart memory alloy (SMA) wires attached to the strip surface. When activated, the SMA wire shortens its length and causes the strip to bend in the direction of the minimum bending stiffness due to the eccentricity between the wire's and the strip's centerline (Fig. \ref{fig:StateSwitch}d). Magnetic actuation provides another solution for generating actuation curvatures by introducing magnetic polarities to the actuated segments \cite{zhao2019mechanics}. The usage of these actuation elements will require the inclusion of the elastic energy of the SMA wire or the magnetic potential, for establishing a full mechanics model.

\section*{Discussion}

Through precision tabletop models and numerical predictions from anistropic rod theory, we demonstrate that introducing radial creases to annular strips enriches their nonlinear mechanics by creating generic bistability and various folding patterns, depending on geometric parameters such as the overcurvature and the crease pattern. Folding of a large closed structure into smaller multiply covered loops is common in slender structures, such as rods/strips \cite{manning2001stability,audoly2015buckling,mouthuy2012overcurvature,starostin2022forceless}, curved folding \cite{dias2014non}, elastic rod networks \cite{yu2021numerical}, and ring origami \cite{wu2021ring}. Different from previous results, we have demonstrated tunable looping patterns in creased annuli. Our numerical modeling is based on a continuous description of creases through a novel regularized Dirac delta function $\Delta$, which captures the localized curvature at creases. We further show that by adding a rest curvature in the direction of the minimum bending stiffness of thin strips, dynamic switches between different states of creased annuli can be achieved. Our framework is convenient for designing and further optimizing the actuation scheme.

In SI Appendix, section 5, we show that as long as creases are sharp (i.e. the extension of the crease $<<$ the length of its adjacent facets), any RDDF could be used to study creased annuli without causing notable differences to its
nonlinear mechanics. The $\Delta$ function proposed in this work could be used to describe other types of material and geometric discontinuities, such as the jump of cross sections in stepped beams \cite{yang2004buckling} and the jump of rest curvatures in serpentine strips \cite{zhang2013buckling}. An example of applying our framework to address discontinuous cross sections and nonlinear material behaviors in the formation of creases is included in SI Appendix, section 6. In SI Appendix, section 7, we further show that a regularized Heaviside function can describe the geometry of 2D surfaces with discontinuities, which provides opportunities for facilitating the mechanics modeling and design of metasheets \cite{liu2023snap,faber2020dome}.

\section*{Materials and Methods}

\noindent{\bf{Fabrication of elastic creased annuli.}} To obtain elastic creased annuli with target geometric parameters, we first make stress-free creased annuli through annealing, with its geometry prescribed by the surface of regular pyramids. Then we cut the annealed annuli and insert/remove flat annular arcs to adjust overcurvature. Based on trial and error, we choose appropriate radii $r_c$ for the tabletop models such that they are not too large and do not suffer significantly from gravity; at the same time, they are not too small and do not cause apparent plastic deformations to the materials. The relation between the geometry of the pyramids and the geometry of creased annuli is detailed in Supplementary Material \ref{appse:FabricationCreaseAnnuli}. \\

\noindent{\bf{Numerical implementation in continuation package AUTO 07P.}} 
Together with anisotropic rod theory, our continuous modeling of creased annuli is implemented as a two-point boundary value problem in AUTO 07P to conduct numerical continuation. We use uniform mesh setting in AUTO to make sure that crease regions have enough meshes. The start solution, boundary conditions, and continuation steps are detailed in Supplementary Material \ref{appse:rodmodel}. \\

\noindent{\bf{Nonlinear stability of creased annuli.}} 
Thanks to our description of creased annuli as a single continuous piece, we are able to determine the stability of equilibria through the standard conjugate point test. After obtaining equilibria through numerical continuation, we first solve an initial value problem with the forces and moments being the initial value at $s=0$, then stability is determined by examining if conjugate points exist in the interval $s \in (0,l]$. Details are documented in Supplementary Material \ref{appse:rodmodel}. \\

\noindent{\bf{Finite element modeling.}} 
Finite element analysis is conducted in the commercial software ABAQUS/Standard and the results are validated against the predictions from our theoretical framework.
We choose Timoshenko beams and set the crease length to $10t$, matching with the crease size implemented in our theoretical framework. The feature ``Nlgeom“ in ABAQUS is turned on to account for geometric nonlinearity.
Temperature gradients are applied along the thickness (only in the crease regions) and the width of the cross section to introduce the crease angle and the overcurvature to the strip, respectively. Young's Modulus and Poisson's ratio are set to 1000 MPa and 0.33, respectively.

\section*{Acknowledgments}

We are grateful to Lauren Dreier, Th\'{e}o Jules, and Andy Borum for useful discussions. TY thanks Andy Borum for sharing the stability test code of anisotropic rod. We are grateful to Lucia Stein-Montalvo for proofreading the manuscript. TY is supported by U.S. National Science Foundation grant CMMI-2122269. FM is supported by Princeton Global Collaborative Network ROBELARCH.

\bibliographystyle{unsrt}
\bibliography{creasedannuli}

\clearpage
\newpage

\begin{center}
	\textbf{\large Supplemental Information} \\
	\large Continuous modeling of creased annuli with tunable bistable and looping behaviors 
\end{center}
\setcounter{equation}{0}
\setcounter{figure}{0}
\setcounter{table}{0}
\setcounter{page}{1}
\setcounter{section}{0}
\makeatletter
\renewcommand{\theequation}{S\arabic{equation}}
\renewcommand{\thefigure}{S\arabic{figure}}
\renewcommand{\thetable}{S\arabic{table}}
\renewcommand{\thesection}{S\Roman{section}}
\renewcommand{\bibnumfmt}[1]{[#1]}
\renewcommand{\citenumfont}[1]{#1}

\section{Crease profiles and errors of the crease angle}\label{appse:AngleError}
The $\Delta$ function used to describe the rest curvature of creased strips leads to creases with certain profiles and errors of the crease angle.
Here, we explore the possible crease profiles and discuss the resulting errors. We focus on a single crease whose rest curvature can be described as,

\begin{equation} \label{eq:singlecrease}
\begin{aligned}
\kappa_0&= \frac{\pi - \gamma} {2 (l_{e} -l_{b} ) } \left[ \tanh \left(\frac{s-l_{b} }{C} \right)-\tanh\left(\frac{s-l_{e} } {C} \right)\right] = (\pi - \gamma) \Delta_C^{(l_b,l_e)} \,, \\
\end{aligned}
\end{equation}

with $\gamma$ corresponding to the target crease angle. $(4C+l_e-l_b)$ and $[l_b-2C, l_e+2C]$ represent the nominal crease length and crease region, respectively. We will show that most of the crease angle is formed in the crease region. First, the total area below the $\Delta$ function is always unity

\begin{equation} \label{appeq:deltaintegral}
\begin{aligned}
\int _{-\infty} ^ {\infty} \Delta^{(l_b, l_e)}_{C} ds =\frac{C}{2 (l_e - l_b)} \left[\ln \left( \cosh \left(\frac{s-l_b}{C} \right) \right)-\ln \left( \cosh \left(\frac{s-l_e}{C} \right) \right) \right] \Bigm\vert  _{-\infty} ^ {\infty}  =1 .
\end{aligned}
\end{equation}

If the rod is infinitely long, we can use Equation \ref{eq:singlecrease} to achieve an exact crease angle $\gamma$. However, the length of a realistic rod is always finite, and thus Equation \ref{eq:singlecrease} leads to a crease with the crease angle always bigger than the target angle $\gamma$, because the total turning angle of the tangent is always smaller than the desired angle $(\pi-\gamma)$. 
In addition, with $(l_e - l_b) \to 0$, $\Delta^{(l_b, l_e)}_{C}$ approaches $\frac{1}{ 2C \cosh^2 \left(\frac{s-l_b}{C} \right) }$, which corresponds to Jules et al.'s description of a single crease in the \emph{elastica} \cite{jules19local}.

\begin{figure}[h!]
	\centering
	\includegraphics[width=0.7\textwidth]{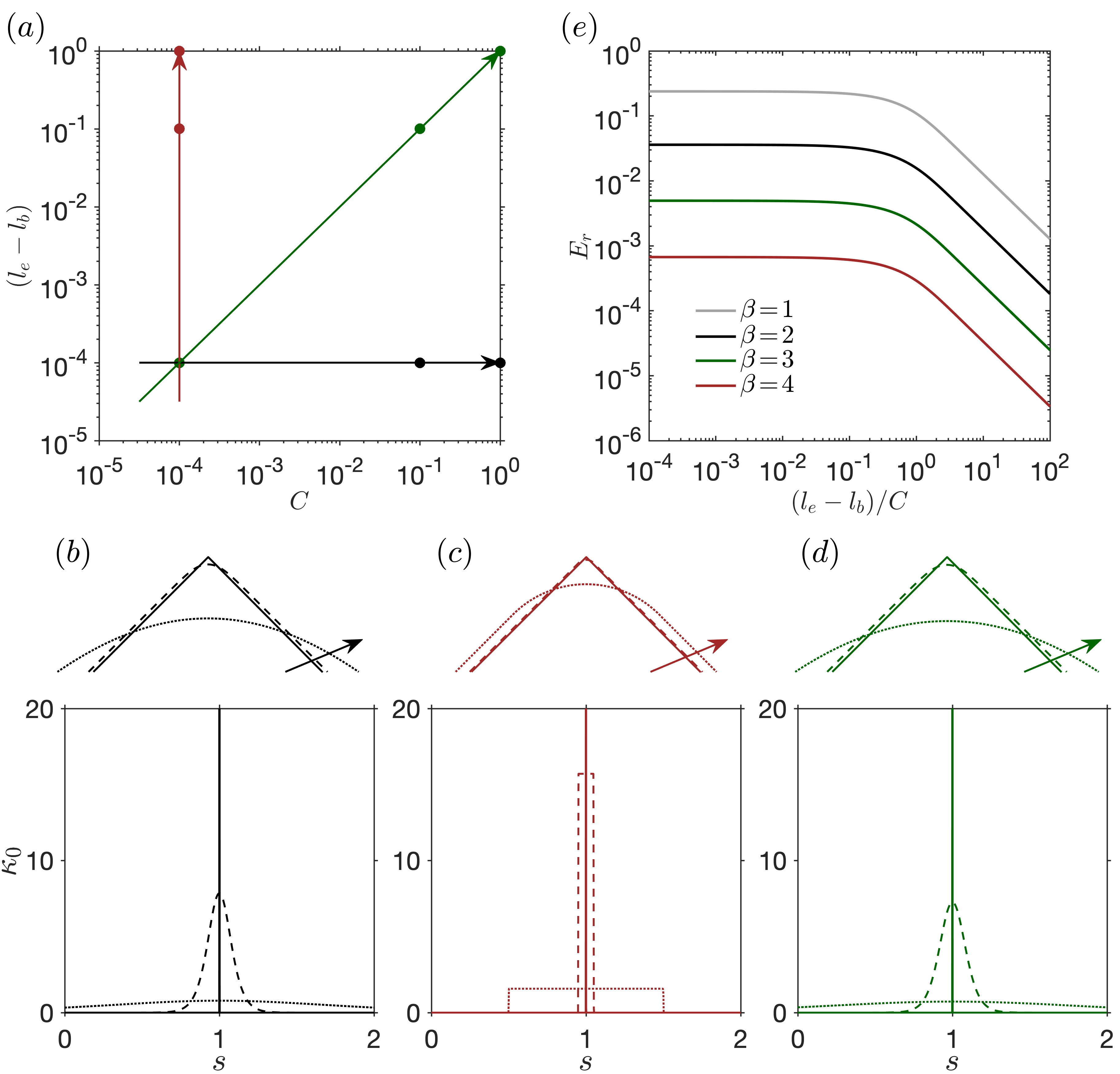}
	\caption{Crease profiles and errors of crease angle with different geometric parameters $C$ and $(l_e-l_b)$. (a) The three arrows represent different regions in $C$ versus $(l_e-l_b)$ plane. (b-d) Crease profiles and their curvature distributions, with geometric parameters corresponding to the circles in (a) and the arrows following these in (a). (e) The error $E_r$ (defined in Equation \eqref{appeq:angleerror}) with different $\beta$.}\label{fig:DeltaCreaseProfile}
\end{figure}

Without the loss of generality, we fix $\frac{l_b + l_e}{2}$ to 1, i.e. the crease is centered at $s=1$. The only remaining parameters affecting the crease profiles are $(l_e-l_b)$ and $C$. 
Figure \ref{fig:DeltaCreaseProfile}(b)-(d) display the crease profiles and rest curvatures with different parameter settings, corresponding to the colored arrows in Figure \ref{fig:DeltaCreaseProfile}(a). We have assumed the length of the rod to be 2 (i.e. $s \in [0,2]$) and crease profiles are obtained by integrating the rest curvature $\kappa_0$. 

Following the black arrow, we fix $(l_e - l_b)$ to a small number $10^{-4}$ and vary $C$ in the range $[10^{-4}, 1]$. With $C=10^{-4}$, the crease is extremely sharp which could be seen through the spike of the curvature. Increasing $C$ makes the crease blunt and $C \!=\! 1$ leads to a shallow arc that does not form the full crease angle. Following the brown arrow, $C$ is fixed to a small number $10^{-4}$ with the increase of $(l_e - l_b)$. In this case, the creases are uniformly bent when $(l_e - l_b) >> C$, e.g. the dashed and dotted renderings in Figure \ref{fig:DeltaCreaseProfile}(c). Following the green arrow, we increase $(l_e - l_b)$ and $C$ simultaneously, resulting in creases similar to those in Figure \ref{fig:DeltaCreaseProfile}(b). The two shallow dotted arcs in Figures \ref{fig:DeltaCreaseProfile}(b) and \ref{fig:DeltaCreaseProfile}(d) do not fully form the crease angle, because the length of the rod is less than the nominal crease length $( 4C+l_e-l_b )$. For other cases, the resulting crease angle is very close to the target angle $\gamma=\pi /2$.

Next, we quantify the errors by calculating the deviation of the integral of $\Delta^{(l_b, l_e)}_{C}$ from unity in the region $[l_b-\beta C,l_e+\beta C]$,

\begin{equation} \label{appeq:angleerror}
\begin{aligned}
E_r=1- \int _{l_b-
	\beta C} ^ {l_e + \beta C} \Delta^{(l_b, l_e)}_{C} ds = 1- \frac{C}{l_e -l_b} \ln \frac{ \cosh (\frac{l_e -l_b}{C} + \beta )}{\cosh \beta}  \,,
\end{aligned}
\end{equation}

which shows that $E_r$ only depends on the ratio $(l_e - l_b)/C$. 
Figure \ref{fig:DeltaCreaseProfile}e reports $E_r$ as a function of $(l_e - l_b)/C$ with different choices of $\beta$. When $(l_e - l_b)/C \! \to \!\! \infty$, $E_r \! \to \!\! 0$ no matter what the value of $\beta$. 
When $(l_e - l_b)=0$, $E_r$ reaches a maximal value of $(1-\tanh \beta)$.
For $\beta=2$, the maximal error is $3.6 \%$. In other words, in the nominal crease region $[l_b-2 C,l_e+2 C]$, at least $96.4 \%$ of the crease angle has been formed.
Increasing $\beta$ reduces the error quickly. For $\beta=4$, $E_r$ is less than $6.71 \times 10^{-4}$.
We conclude that as long as the extremities of the rod are outside the nominal crease region, we will obtain an accurate crease angle. 

\newpage

\section{NYC-skyline function} \label{appse:NYCskyline}

\begin{figure}[h!] 
	\centering
	\includegraphics[width=0.4\textwidth]{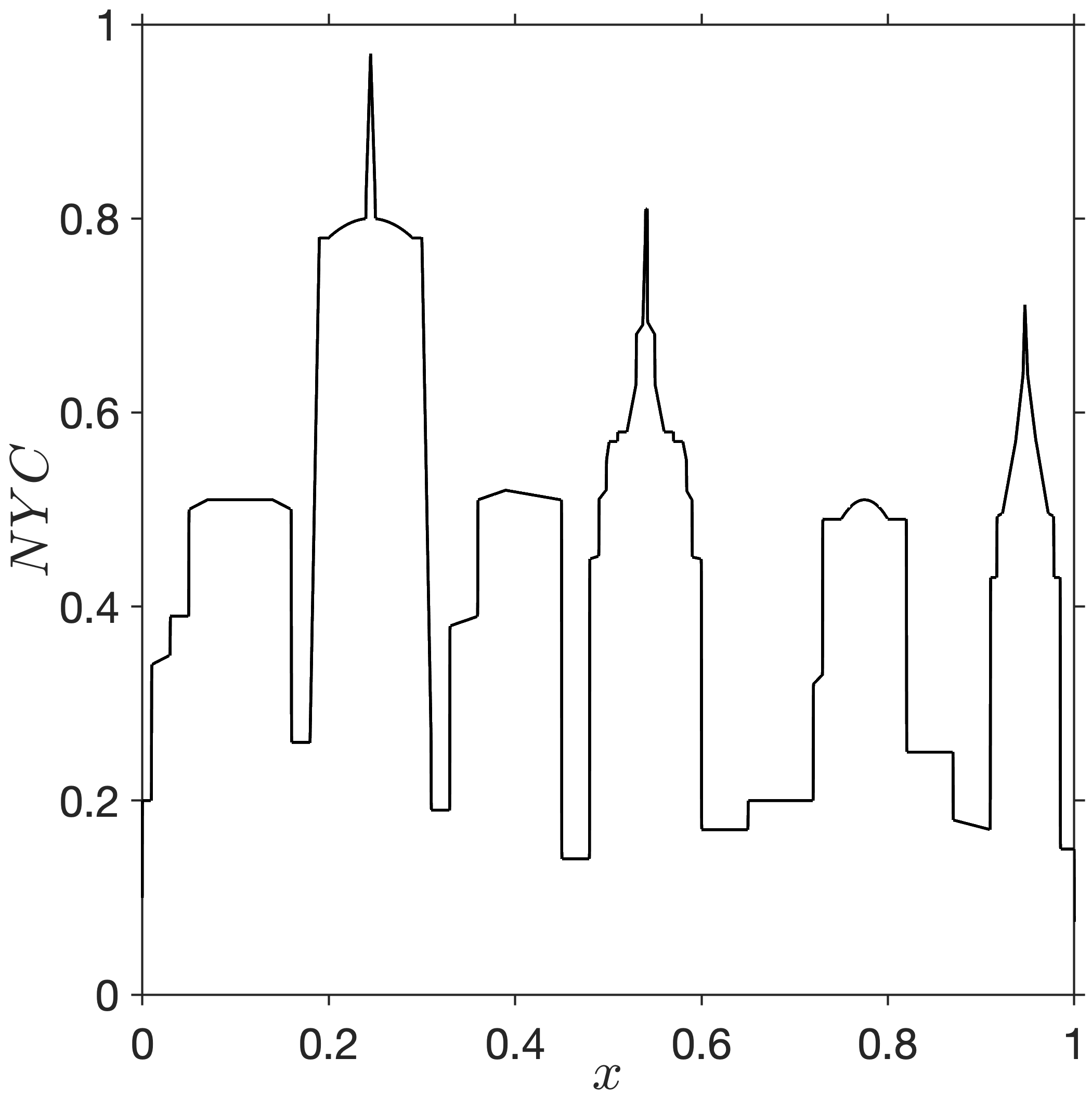}
	\caption{NYC skyline, described by Equation \eqref{eq:NYCskyline} as a continuous piece.} \label{appfig:NYCskyline}
\end{figure}

To characterize the NYC skyline in Figure \ref{appfig:NYCskyline} as a continuous function with $C^{\infty}$, we use the boxcar feature of the $\Delta$ function in Equation \eqref{eq:Deltafunc}. 
Figure \ref{appfig:NYCskyline} describes part of the New York City skyline through the following expression

\begin{equation}\label{eq:NYCskyline}
NYC(x)=\sum_{i=1}^{53} \frac{f_i(x)}{2} \left[ \tanh \left(\frac{x-l _{bi} }{C_i}\right)-\tanh\left(\frac{x-l _{ei} }{C_i}\right)\right] \,,
\end{equation}

where $f_i(x)$ represents each segment of the piecewise continuous profile of the NYC skyline. We set $C_i$ to be a small number $10^{-4}$ for all $i \in [1,53]$ and omit the prefactor $1/(l_{ei} - l_{bi})$, which leads to $\Delta\!=\!1$ in $x \in [l_{bi},l_{ei}]$ and $\Delta\!=\!0$ elsewhere. Here $\Delta$ works like a switch that only turns on $f_i(x)$ in $x \in [l_{bi},l_{ei}]$. Parameters and expressions of $NYC(x)$ are summarized in table \ref{tab:NYCpars}.

\begin{table}[]
	\centering
	\begin{tabular}{|r|lll||r|lll|}\hline
		$i$ & $l_{bi}$ & $l_{ei}$ & $f_i(x)$ &                        $i$ & $l_{bi}$ & $l_{ei}$ & $f_i(x)$ \\ \hline
		$1$ & $0.0$ & $0.01$ & $0.2$ &                          $28$ & $0.542$ & $0.544$ & $27.89 - 50 s$ \\    
		$2$ & $0.01$ & $0.03$ & $0.335 + 0.5 s$ &               $29$ & $0.544$ & $0.55$ & $1.597 - 1.667 s$ \\
		$3$ & $0.03$ & $0.05$ & $0.39$ &                        $30$ & $0.55$ & $0.56$ & $3.38 - 5. s$ \\          
		$4$ & $0.05$ & $0.07$ & $0.475 + 0.5 s$ &               $31$ & $0.56$ & $0.57$ & $0.58$ \\                 
		$5$ & $0.07$ & $0.14$ & $0.51$ &                        $32$ & $0.57$ & $0.58$ & $0.57$ \\                 
		$6$ & $0.14$ & $0.16$ & $0.58 - 0.5 s$ &                $33$ & $0.58$ & $0.584$ & $3.47 - 5 s$ \\          
		$7$ & $0.16$ & $0.18$ & $0.26$ &                        $34$ & $0.584$ & $0.59$ & $1.493 - 1.667 s$ \\     
		$8$ & $0.18$ & $0.19$ & $-9.1 + 52 s$ &                 $35$ & $0.59$ & $0.6$ & $0.569 - 0.2 s$ \\         
		$9$ & $0.19$ & $0.30$ & $0.78$ &                        $36$ & $0.6$ & $0.65$ & $0.17$ \\                  
		$10$ & $0.20$ & $0.29$ & $0.02 - 9.877 (s-0.245)^2$ &   $37$ & $0.65$ & $0.72$ & $0.2$ \\                  
		$11$ & $0.24$ & $0.245$ & $-7.18 + 30 s$ &              $38$ & $0.72$ & $0.73$ & $-0.4 + 1. s$ \\          
		$12$ & $0.245$ & $0.25$ & $7.52 - 30 s$ &               $39$ & $0.73$ & $0.82$ & $0.49$ \\                 
		$13$ & $0.3$ & $0.31$ & $16.38 - 52 s$ &                $40$ & $0.75$ & $0.8$ & $0.02 - 32 (s -0.775)^2$ \\
		$14$ & $0.31$ & $0.33$ & $0.19$ &                       $41$ & $0.82$ & $0.87$ & $0.25$ \\                 
		$15$ & $0.33$ & $0.36$ & $0.27 + 0.333 s$ &             $42$ & $0.87$ & $0.91$ & $0.3975 - 0.25 s$ \\      
		$16$ & $0.36$ & $0.39$ & $0.39 + 0.333 s$ &             $43$ & $0.91$ & $0.917$ & $0.43$ \\                
		$17$ & $0.39$ & $0.45$ & $0.585 - 0.167 s$ &            $44$ & $0.917$ & $0.923$ & $0.0345 + 0.5 s$ \\     
		$18$ & $0.45$ & $0.48$ & $0.14$ &                       $45$ & $0.923$ & $0.937$ & $-4.383 + 5.286 s$ \\   
		$19$ & $0.48$ & $0.49$ & $0.305 + 0.3 s$ &              $46$ & $0.937$ & $0.945$ & $-7.512 + 8.625 s$ \\   
		$20$ & $0.49$ & $0.498$ & $-0.102 + 1.25 s$ &           $47$ & $0.945$ & $0.947$ & $-33.381 + 36 s$ \\     
		$21$ & $0.498$ & $0.501$ & $-2.77 + 6.667 s$ &          $48$ & $0.947$ & $0.95$ & $23.439 - 24 s$ \\       
		$22$ & $0.501$ & $0.51$ & $0.57$ &                      $49$ & $0.95$ & $0.959$ & $7.922 - 7.667 s$ \\     
		$23$ & $0.51$ & $0.52$ & $0.58$ &                       $50$ & $0.959$ & $0.972$ & $6.029 - 5.692 s$ \\    
		$24$ & $0.52$ & $0.53$ & $-2.02 + 5 s$ &                $51$ & $0.972$ & $0.978$ & $0.982 - 0.5 s$ \\      
		$25$ & $0.53$ & $0.537$ & $-0.077 + 1.429 s$ &          $52$ & $0.978$ & $0.985$ & $0.43$ \\               
		$26$ & $0.537$ & $0.54$ & $-17.21 + 33.333 s$ &         $53$ & $0.985$ & $1.0$ & $0.15$ \\                 
		$27$ & $0.54$ & $0.542$ & $0.81$                             &         &       &        \\
		\hline
	\end{tabular}
	\caption{Parameters and expressions used in the NYC skyline function in Equation \ref{eq:NYCskyline}. $C_i$ is fixed to $0.0001$}
	\label{tab:NYCpars}
\end{table}

\newpage

\section{Fabrication of creased annular strips} \label{appse:FabricationCreaseAnnuli} 
\begin{figure}[h!]
	\centering
	\includegraphics[width=0.6\textwidth]{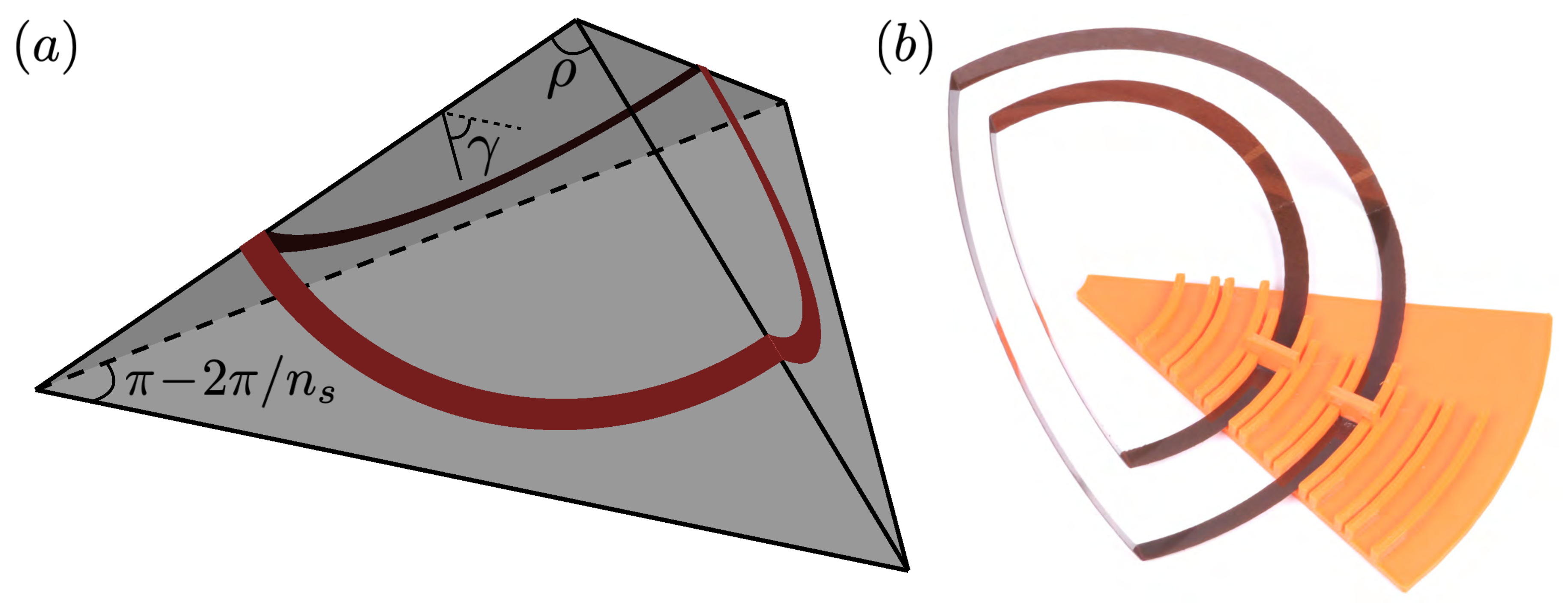}
	\caption{3D printed molds are used to construct precision tabletop models. (a) Geometry of a regular pyramid with a triangular base used to prescribe the crease angle of a creased annular strip to $\gamma$. (b) 3D printed circular grooves used for precise alignment in closing creased annuli with super glue.} \label{fig:PyramidGeometry}
\end{figure}

We 3D print regular pyramids to anneal creases, which fix the crease angle to $\gamma$, eliminate the residual stresses, and result in elastic creases \cite{klett2018paleo,sargent2019heat,jules19local}. Figure \ref{fig:PyramidGeometry}a shows an example of the geometry of a regular pyramid with a triangular base and a superimposed creased annular strip with three creases. The geometry of the pyramid is fully determined by the ridge angle $\gamma$ (i.e. the dihedral angle between two adjacent lateral faces), which prescribes the crease angle of creased annuli. Since $\gamma$ is always larger than the internal angle of the polygonal base $(\pi - 2\pi /n_s)$, a pyramid with a polygon base of $n_s$ sides can only be used for annealing creases with an angle larger than $(\pi - 2\pi /n_s)$.  

With a prescribed ridge angle $\gamma$, the vertex angle of a lateral face $\rho$ can be obtained as

\begin{equation}\label{eq:pyramidgeometry}
\begin{aligned}
\rho &=2 \sin^{-1} \sqrt{ \frac{\cos \gamma +\cos \frac{2\pi}{n_s}}{ \cos \gamma -1} }  \,, 
\end{aligned}
\end{equation}

In order to match the pyramid surface exactly, we laser cut annular strip with an effective length $\rho r_c n_s$ (excluding the gluing length of 2 mm), where $r_c$ corresponds to the radius of the centerline. We use super glue to close the strip with the thickness of the gluing regions being slightly engraved in the cutting process. 
To achieve precise gluing of the two ends of annular strips, we 3D print a gadget consisting of a series of circular grooves, through which the two ends of the annular strip are pushed inside the groove to achieve good alignment (See Figure \ref{fig:PyramidGeometry}b). Even though the joint area almost doubles the thickness of the strip, its effects on the nonlinear mechanics of creased annuli are negligible due to its small length compared with the length of the annuli. 
After being annealed, the overcurvature of the stress-free creased annuli is $O_c=\rho n_s/(2\pi)$. We further cut the annealed strip and insert/remove annular arcs to adjust its overcurvature $O_c$. The inserted flat annular arcs are annealed together with the creased annular strips to make sure all the components of creased annuli have the same material properties.

Theoretically, the shape of creased annuli with a fixed cross section of the strip does not depend on the size of the experimental models, as long as the strip is slender (i.e. $L >> w$). However, models with a larger length will be more flexible and thus suffer more from gravity. On the other hand, smaller models have larger curvatures could experience plastic deformations. The sizes of the physical models (determined by radius $r_{c}$) are selected by trials and errors, such that the tabletop models do not suffer significantly from gravity, and at the same time, do not cause apparent plastic deformations to the strips. With the strip cross section fixed to $w \!=\! 5.08$ mm and $t \!=\! 0.254$ mm, we used $r_c \!=\! 40$ mm for $O_c \!=\! 3$, $r_c \!=\! 50$ mm for $O_c \!=\! 1.5$, and $r_c \!=\! 70$ mm for $O_c \!=\! 0.7$ in our tabletop models.

\section{anisotropic rod model and  nonlinear stability of equilibria} \label{appse:rodmodel} 

Anisotropic rod theory is normally used to model slender rods with the two dimensions of the cross section being in the same order of magnitude, e.g. $w \! \sim \! t$ for a strip with a rectangular cross section. Recent studies have shown that anisotropic rod model works quantitatively well for capturing the nonlinear mechanics of thin strips with $w/t$ up to $O (10)$ \cite{riccobelli2021rods,moulton2018stable,yu2019bifurcations}. Here, we use anisotropic rod theory to solve the nonlinear mechanics of creased annuli. 
Throughout this study, we fix the aspect ratio of the cross section $w/t$ to 20. 

\begin{figure}[h!]
	\centering
	\includegraphics[width=0.75\textwidth]{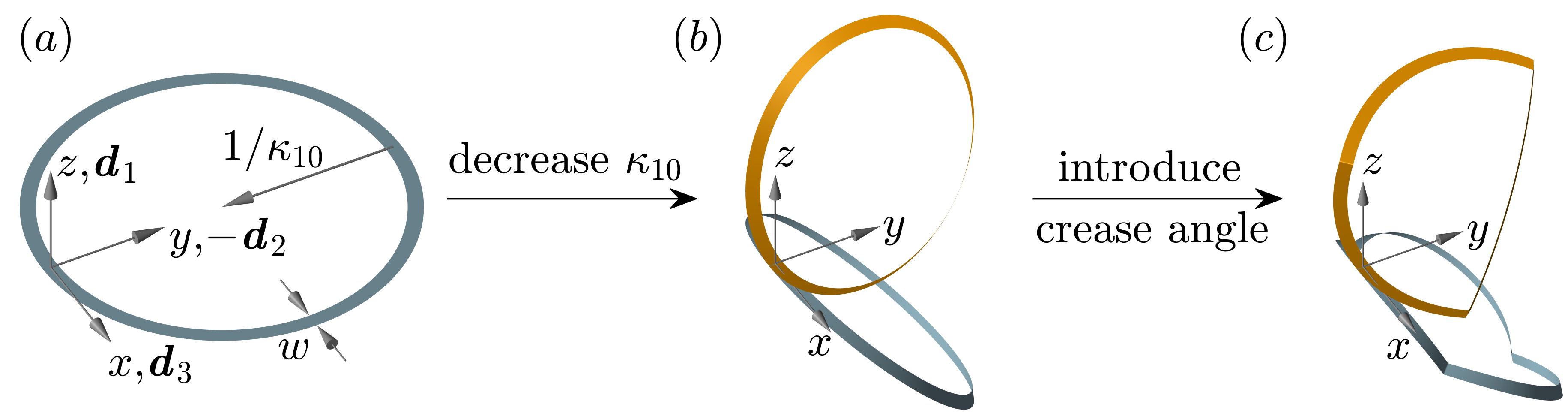}
	\caption{Several continuation steps used to obtain the folded state and inverted state. (a) Start solution: a stress-free annular strip. (b) A pair of bifurcated conical shapes. (c) Folded (orange) and inverted states (blue).} \label{fig:bigonKinematics}
\end{figure}

A local orthonormal right-handed material frame $(\bm{d}_1,\bm{d}_2,\bm{d}_3)$ is attached to the centerline of the strip, with $\bm{d}_3$ corresponding to the local tangent, and $\bm{d}_1$ and $\bm{d}_2$ aligned with the width $w$ and thickness $t$, respectively (see Figure \ref{fig:bigonKinematics}a). The kinematics of the material frame can be described as $\bm{d}_i'=\bm{\omega} \times \bm{d}_i$, with $\bm{\omega}=\kappa_{1} \bm{d}_1 + \kappa_{2} \bm{d}_2 +\tau \bm{d}_3$ and $\kappa_{1}, \kappa_{2}$ and $\tau$ corresponding to two bending curvatures and the twist, respectively.
The internal forces and moments can be further resolved on the material frame as $\bm{N}=N_1 \bm{d_1} +N_2 \bm{d_2} +N_3 \bm{d_3}$ and $\bm{M}=M_1 \bm{d_1} +M_2 \bm{d_2} +M_3 \bm{d_3}$. Substituting the resolved forces and moments and the kinematics of the material frame into the equilibrium Equations (i.e. Eq. (3) in the main text), we have 

\begin{equation}\label{eq:F&Mequilibrium1} 
\begin{aligned}
N_1'-N_2 \tau+N_3 \kappa_2=0 \, , N_2'+N_1 \tau-N_3 \kappa_1=0 \, , N_3'+N_2 \kappa_1-N_1 \kappa_2 =0 \, , \\
M_1'-M_2 \tau-N_2+M_3 \kappa_2 =0 \, , M_2'+M_1 \tau-M_3 \kappa_1+N_1 =0 \, , M_3'+M_2 \kappa_1-M_1 \kappa_2 =0 \, .
\end{aligned}
\end{equation}

We use linear constitutive laws $M_1 \!=\! EI_1 (\kappa_1 - \kappa_{10})$, $M_2 \!=\! EI_2 (\kappa_2 - \kappa_{20})$, and $M_3 \!=\! GJ \tau$, where $E$ and $G$ are the Young's modulus and shear modulus, respectively. $EI_1$, $EI_2$, and $GJ$ are the two bending rigidities and the torsional rigidity, respectively. $\kappa_{10}$ and $\kappa_{20}$ correspond to the two bending curvatures of the rest configuration. Here, $\kappa_{10} \!=\! 2 \pi O_c/L$ and $\kappa_{20} \!=\! \sum_{i=1}^{n_c} (\pi - \gamma_i)   \Delta_{C_i} ^{(l_{bi},l_{ei})} $, correspond to the geodesic curvature of the annular strip and the curvature of creases, respectively.
Substituting the constitutive laws into \eqref{eq:F&Mequilibrium1}, dividing the two sides of each equation by $GJ$, and defining the stiffness ratios $a \!=\! EI_1/(GJ)$ and $b \!= \! EI_2/(GJ)$, we have

\begin{equation}\label{appeq:normF&M} 
\begin{aligned}
& N_1'  = (N_2 \tau -N_3 \kappa_2) l \, , N_2'  = (-N_1 \tau + N_3 \kappa_1 ) l \, , N_3'  =  (-N_2 \kappa_1 + N_1 \kappa_2) l \, , \\
& a (\kappa_1 ' - \kappa_{10}' ) = (b (\kappa_2 - \kappa_{20}) \tau - \tau \kappa_2 + N_2) l \, , \\
& b (\kappa_2 ' - \kappa_{20} ')  = (-a (\kappa_1 - \kappa_{10}) \tau + \tau \kappa_1  - N_1) l \, , \\
& \tau'  = (-b (\kappa_2 - \kappa_{20} )  \kappa_1 + a (\kappa_1 - \kappa_{10}) \kappa_2) l  \,, \\
& q'_1 =\left( \tfrac{1}{2}(-q_2 \tau +q_3 \kappa_2 -q_4 \kappa_1)+ \mu q_1 \right) l \,,  q'_2=\left( \tfrac{1}{2}(q_1 \tau + q_4 \kappa_2 +q_3 \kappa_1) + \mu q_2 \right) l  \,, \\
& q'_3 =\left( \tfrac{1}{2}(q_4 \tau - q_1 \kappa_2 - q_2 \kappa_1) + \mu q_3 \right) l \,, q'_4=\left( \tfrac{1}{2}(-q_3 \tau -q_2 \kappa_2 +q_1 \kappa_1) + \mu q_4 \right) l  \,, \\
& x' =2(q_1^2 + q_2^2-\tfrac{1}{2}) l \,, \; y' =2(q_2 q_3 + q_1 q_4) l \,,\; z ' =2(q_2 q_4 - q_1 q_3) l \,, s'=l \,, \\
\end{aligned}
\end{equation}

where a prime denotes an $\bar{s}$ derivative ($\bar{s} \!=\! s/l \in [0,1]$ ) and $(q_1,q_2,q_3,q_4)$ corresponds to the unit quaternions. In this study, the arc length $s$ explicitly enters the equation through $\kappa_{20}$ and the final ODE $s' \!=\! l$ turns the system into a standard boundary value problem. Following Healey and Mehta \cite{healey2006straightforward}, the dummy parameter $\mu$ in \eqref{appeq:normF&M} allows a consistent prescription of boundary conditions for quaternions. In numerical continuation, we treat $\mu$ as a free parameter and keep monitoring its value, which should be numerically zero \cite{healey2006straightforward}. In our work, $\mu$ is found to be in the order of $10^{-14}$.

For rods with a rectangular cross section composed of an elastically isotropic material, the bending and twisting stiffness are \cite{timoshenko1951theory},

\begin{equation}\label{eq:inertiaofmoment} 
\begin{aligned} 
EI_1=\tfrac{1}{12} E w^3 t\,,\;  EI_2=\tfrac{1}{12} E w t^3 \,,\; 
GJ=\lambda G  w t^3=\lambda \frac{E}{2(1+\nu)} w t^3 \, ,
\end{aligned}
\end{equation}

which leads to 
\begin{equation}\label{eq:coefficients} 
a=\frac{EI_1}{GJ}=\frac{(1+\nu)}{6\lambda} \left(\frac{w}{t} \right)^2 \,, b=\frac{EI_2}{GJ}=\frac{(1+\nu)}{6\lambda}  \, .
\end{equation}

Here $\nu$ is the Poisson's ratio, which is set to 0.33 in this study. $\lambda$ depends on the aspect ratio of the cross section. For $w/t \!=\! 20$, we have $\lambda \!=\! 0.3228$ \cite{timoshenko1951theory}. Our formulation shows that material properties have minor influences on the nonlinear mechanics of creased annuli. First, the Young's Modulus does not appear in the normalized bending rigidities. Second, it is known that Poisson's ratio has minimal effects on the mechanics of anisotropic rods (6). We conclude that the nonlinear mechanics of creased annuli are mainly determined by the geometric parameters  (i.e. the number of creases $n_c$, the crease angle $\gamma$, the overcurvature $O_c$, and the aspect ratio of the cross section of the strip $w/t$).

For the creased annuli, we clamp the middle of one segment at the origin of a Cartesian coordinate, with $\bm{d_1}$, $\bm{d_2}$, and $\bm{d_3}$ aligned to $z$, $-y$, and $x$ direction, respectively (Figure \ref{fig:bigonKinematics}a). The boundary conditions can be summarized as

\begin{equation}\label{appeq:boundaryconditions} 
\begin{aligned}
&x(0) =0 \,, y(0) =0\,, z(0) =0\,, q_1(0)=1 \,, q_2(0)=0\,, q_3(0)=0\,, q_4(0)=0 \,, \\
&x(1) =0\,, y(1) =0\,, z(1) =0\,, q_1(1)=-1 \,, q_2(1)=0\,, q_3(1)=0\,, q_4(1)=0 \,, s(0) =0 \,.
\end{aligned}
\end{equation}  

\eqref{appeq:normF&M} and \eqref{appeq:boundaryconditions} make a well-posed two point boundary value problem with fifteen unknowns $N_1$, $N_2$, $N_3$, $\kappa_{1}$, $\kappa_{2}$, $\tau$, $q_1$, $q_2$, $q_3$, $q_4$, $x$, $y$, $z$, $s$, and $\mu$. To solve the boundary value problem, we conduct numerical continuation through AUTO 07P, which uses
orthogonal collocation and pseudo-arclength continuation to trace the solutions and detect bifurcations and limit points. We use a stress-free planar annulus in the $y-x$ plane as a starting point for continuation. The exact solution of the planar circle can be summarized as

\begin{equation}\label{appeq:startsols} 
\begin{aligned}
&N_1 =0 \,, N_2 =0\,, N_3 =0\,, \kappa_{1}=2 \pi/l \,, \kappa_{2}=0 \,, \tau=0 \,, q_1=\cos (\pi \bar{s}) \,, q_2=0\,, q_3=0\,, q_4=\sin (\pi \bar{s}) \,, \\
&x =\tfrac{l}{2\pi} \sin (2 \pi \bar{s})  \,, y =\tfrac{l}{2\pi} [1 - \cos (2 \pi \bar{s})] \,, z =0\,, s =l \bar{s} \,; \,\, \,\,\,\, \mu=0 \,, \gamma=\pi \,,O_c=1 \,.
\end{aligned}
\end{equation}

To obtain the folded and inverted state, we first decrease $\kappa_{10}$ by decreasing $O_c$ 
($\kappa_{10} \!=\! 2 \pi O_c/l$), which leads to a pair of stable conical shapes bifurcated from the planar branch. Then we add crease angle to the bifurcated branches to obtain a folded and an inverted state, respectively. Finally, we systematically study how geometric parameters such as overcurvature and number of creases affect the nonlinear mechanics of creased annuli.

To determine the stability of an equilibrium solution $(\kappa_{1}, \kappa_{2}, \tau, N_1 , N_2, N_3)$, we adopted a method from geometric mechanics to test the existence of conjugate point \cite{bretl2014quasi,borum2020helix}, which applies to a single continuous rod with clamp-clamp boundary conditions. We solve the following matrix differential equations \cite{bretl2014quasi,borum2020helix},

\begin{equation}\label{appeq:IVP} 
\begin{aligned}
D'=FD \,, P'=QD + HP \,,\\
\end{aligned}
\end{equation}

where a prime denotes an $s$ derivative ($s\in[0,l]$ with $l$ normalized to $n_c$). The coefficient matrices $F$, $Q$, and $H$ can be written as \cite{bretl2014quasi,borum2020helix},

\begin{equation}\label{appeq:F2} 
\begin{aligned}
F
&=
\begin{bmatrix} 0 & M_2(\frac{1}{b} - \frac{1}{a}) + \kappa_{20} & M_1 (\frac{1}{b} - \frac{1}{a}) - \kappa_{10} & 0 & 0 & 0 \vspace{4pt} \\
M_2 (1 - \frac{1}{b}) - \kappa_{20} & 0 & M_3 (1 - \frac{1}{b}) & 0 & 0 & 1 \vspace{4pt} \\ 
M_1 (\frac{1}{a} - 1) + \kappa_{10} & M_3 (\frac{1}{a} - 1) & 0  & 0 & -1 & 0 \vspace{4pt} \\
0 & -\frac{N_2}{a} & \frac{N_1}{b} & 0 & \kappa_2 & - \kappa_1 \vspace{4pt} \\
N_2 & 0 & -\frac{N_3}{b} & - \kappa_2 & 0 & \tau \vspace{4pt} \\
-N_1 & \frac{N_3}{a} & 0 & \kappa_1 &- \tau & 0 
\end{bmatrix} 
\end{aligned}
\end{equation}

\begin{equation}\label{appeq:Q} 
\begin{aligned}
Q	
&=
\begin{bmatrix} 1 & 0 & 0 & 0 & 0 & 0 \\
0 & 1/a & 0 & 0 & 0 & 0 \\
0 & 0 & 1/b & 0 & 0 & 0 \\
0 & 0 & 0 & 0 & 0 & 0 \\
0 & 0 & 0 & 0 & 0 & 0 \\
0 & 0 & 0 & 0 & 0 & 0 \\
\end{bmatrix} \,,
H
&=
\begin{bmatrix} 0 & \kappa_2 & -\kappa_1 & 0 & 0 & 0 \\
- \kappa_2 & 0 & \tau & 0 & 0 & 0 \\ 
\kappa_1 & -\tau & 0  & 0 & 0 & 0 \\ 
0 & 0 & 0 & 0 & \kappa_2 & -\kappa_1 \\
0 & 0 & 1 & -\kappa_2 & 0 & \tau \\
0 & -1 & 0 & \kappa_1 & -\tau & 0 
\end{bmatrix} \,,
\end{aligned}
\end{equation}

where $\kappa_1 \!=\! \frac{M_1}{a} + \kappa_{10}$, $\kappa_2 \!=\! \frac{M_2}{b} + \kappa_{20}$, and $\tau \!=\! M_3 $ (notice that torsional stiffness has been normalized to unity). Together with the initial conditions $D(0) \!=\! I_{6 \times 6}$ and $P(0) \!=\! 0_{6 \times 6}$, we obtain an initial value problem that is solved with MATLAB. If the solution of \eqref{appeq:IVP} satisfies $\det(P) \! \ne \! 0$ for all $s \in (0,l]$, then the equilibrium is stable. If $\det(P) \!=\! 0$ for some $s \in (0,l]$, then the equilibrium is unstable. If $\det(P) \! \ne \! 0$ for all $s \in (0,l)$ and  $\det(P(l)) \!=\! 0$, then we cannot conclude the stability and higher order variations of the energy functional need to be considered \cite{bretl2014quasi,borum2020helix}.

\begin{figure}[h!]
	\centering
	\includegraphics[width=0.8\textwidth]{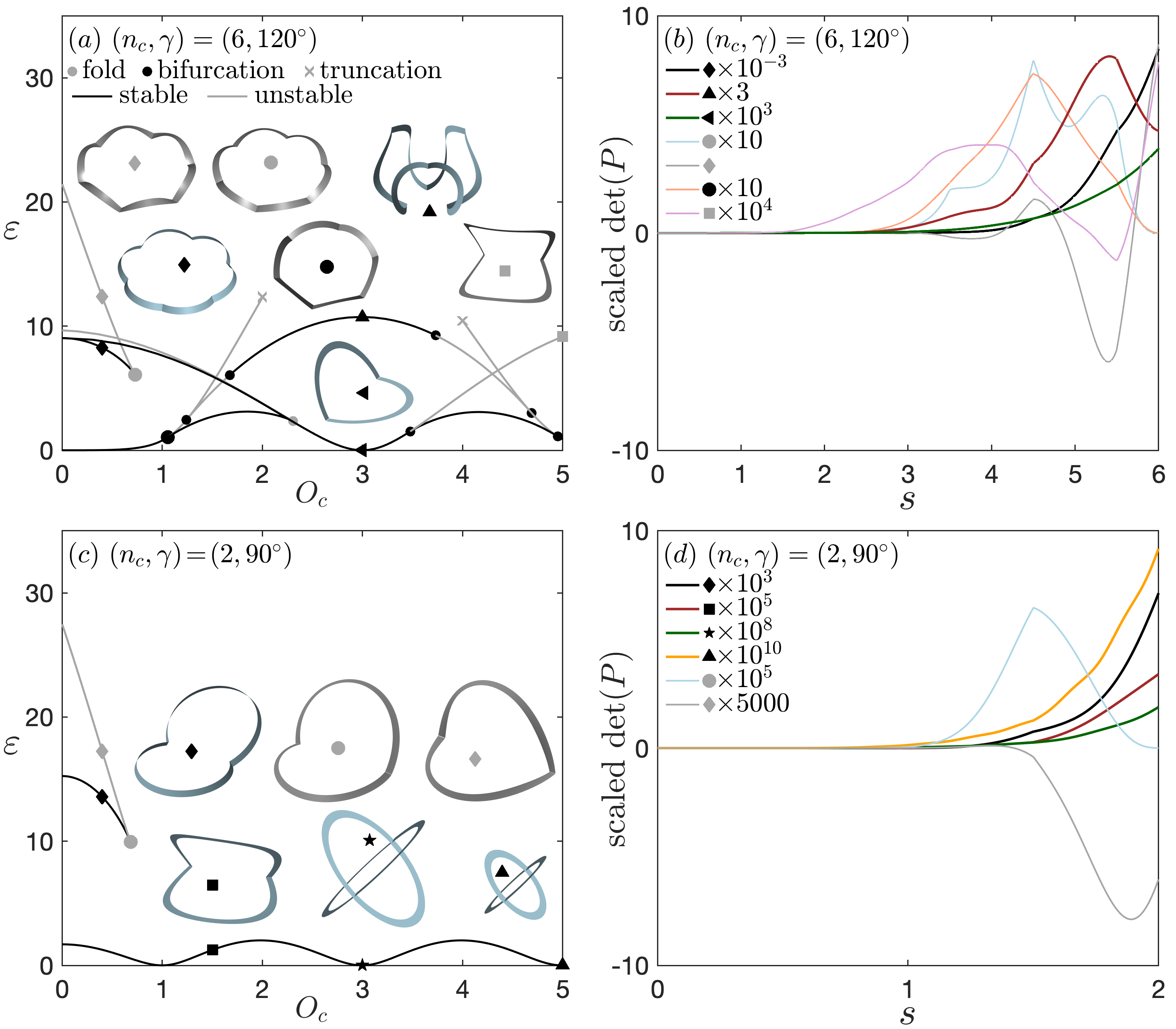}
	\caption{Bifurcation curves (black: stable; grey: unstable ) and examples of conjugate point tests of creased annuli. (a) Solutions with six creases. (b) Conjugate point tests of several solutions in (a). (c) Solutions with two creases. (d) Conjugate point tests of several solutions in (c).}\label{fig:AnnularStability}
\end{figure}

Figure \ref{fig:AnnularStability}(a-b) and \ref{fig:AnnularStability}(c-d) summarized the solutions and conjugate point tests of creased annuli with six creases ($l \!=\! 6$) and two creases ($l \!=\! 2$), respectively. The bifurcation curves are presented in the energy $\varepsilon$ versus overcurvature $O_c$ plane, with $(C_i, l_{ei}-l_{bi})$ fixed to $(0.002,4 \times 10^{-7})$.
Some of the test results $\det(P)$ are multiplied with a factor, which does not affect the stability information. For example, in Figure \ref{fig:AnnularStability}(b), the black curve represents $10^{-3} \times $ the test result of the solution $\blackdiamond$ in \ref{fig:AnnularStability}(a).
In all the tests, unstable solutions contain at least one conjugate point (where the scaled $\det(P)$ crosses zero) and stable solutions are free of conjugate points in $s \in (0,l]$.
Right at a critical point such as a fold or bifurcation point, $\det P(l) \!=\! 0$ and a conjugate point is located exactly at $s \!=\! l$ end. For example, the $\blackcircle$ and $\graycircle$ in Figure \ref{fig:AnnularStability}(a-b) and the $\graycircle$ in Figure \ref{fig:AnnularStability}(c-d).

\section{different regularized Dirac delta functions} \label{appse:DiffDelta}

Here, we use the following RDDFs to describe the local geometry of a crease centered at $s \!=\! 1$

\begin{equation}\label{eq:DiracDelta}
\begin{aligned}
\delta_1 &=\frac{1}{\pi} \frac{C_1}{(s-1)^2+C_1^2} \,, \delta_2 =\frac{1}{ 2C_2 } \frac{1}{ [1+ (\frac{s-1}{C_2})^6] ^{7/6} }\,, \delta_3=\frac{1}{2} (\delta_1 + \delta_2) \,. \\
\end{aligned}
\end{equation}

\begin{figure}[h!]
	\centering
	\includegraphics[width=0.9\textwidth]{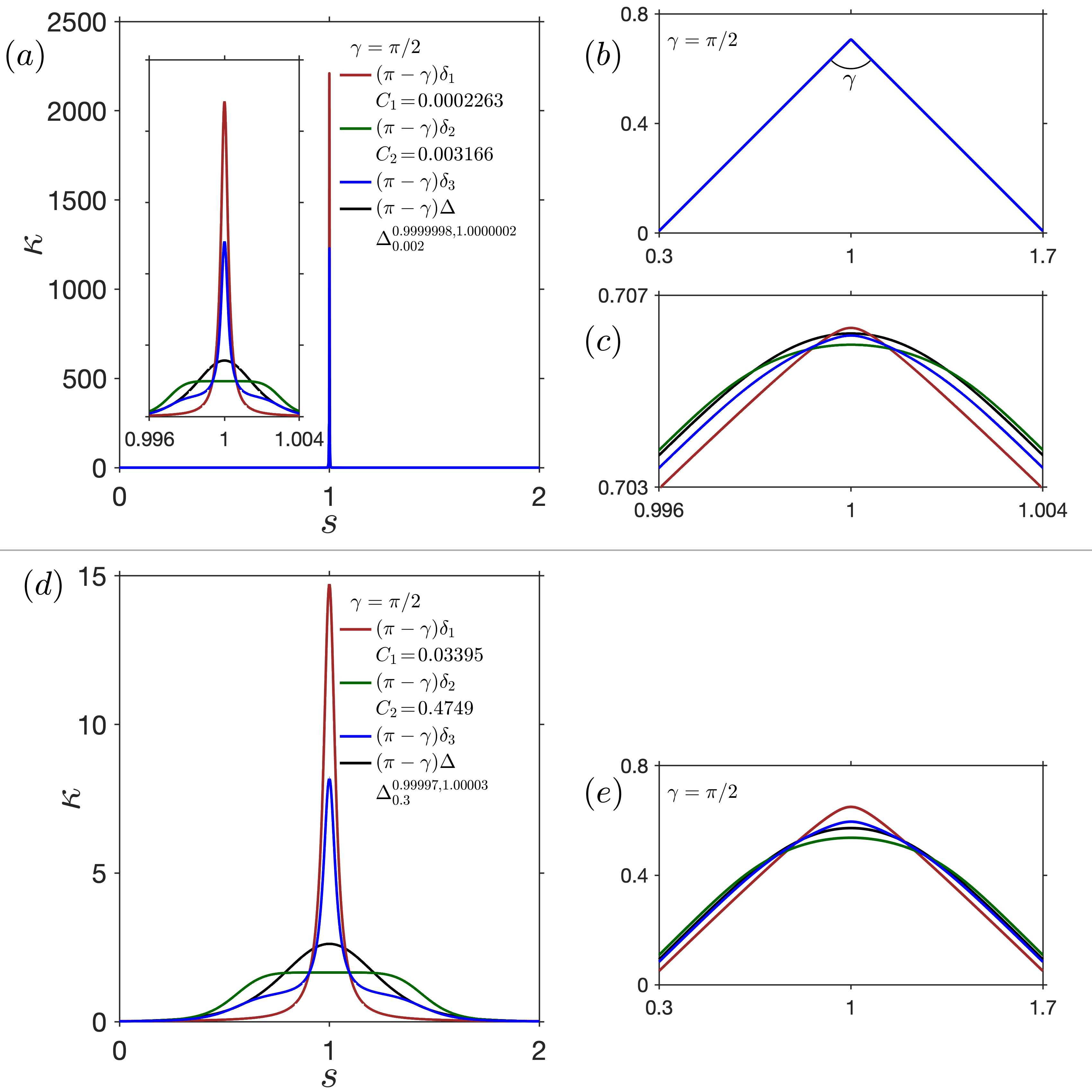}
	\caption{Different RDDFs (Equation \eqref{eq:DiracDelta}) are used to describe the local geometry of a crease centered at $s=1$ with a normalized length 2. (a) Curvature distribution of a sharp crease with $\gamma=\pi/2$. (b) The corresponding rod has no notable differences. (c) A blowup of the local crease region shows significant differences across different RDDFs. (d) Creases that are 150 times the size of (a) are obtained by scaling the geometric parameters accordingly. (e) The resulted rod geometries from (d) contains significant differences.} \label{fig:VariousCrease}
\end{figure}

The rest curvature of a crease with target angle $\gamma$ can be described as $(\pi-\gamma) \delta_i$. Here $C_i$ controls the sharpness of the crease and a perfect crease with $C^0$ continuity is achieved by taking the limit $C_i \to 0$.
$C_i$ is obtained by setting the nominal crease length of $\delta_i$ to be the same with $\Delta$. Specifically, in Figure \ref{fig:VariousCrease}a, we have $ \int _{0.996} ^ {1.004} \delta_i ds=96.4 \%$ for all $i \!=\! 1,2,3$, which is also true for $\Delta$ with $C \!=\! 0.002$ and $(l_e-l_b) \!=\! 2 \times 10^{-4} C$.

Figure \ref{fig:VariousCrease}(a) shows the curvature distribution of different RDDFs with $\gamma$ fixed to $0.5 \pi$. While the curvature clearly contains a similar localized feature around the crease at $s=1$ (shown as spikes with different heights), the inset shows the significant differences of the curvature in the nominal crease region. The crease profiles generated by these $\delta$ functions can be obtained by integrating the curvature, which are shown together in Figure \ref{fig:VariousCrease}(b) for the global shape and Figure \ref{fig:VariousCrease}(c) for the local crease geometry. We remark that even though the local crease geometry varies significantly across different RDDFs, no notable differences are found in the global shapes. Intuitively, we would expect that if the various right-angle rods in Figure \ref{fig:VariousCrease}(b) are subject to the same loading, the nonlinear mechanics responses would be the same, even though the rods may contain very different local deformations around the crease region. 
Figure \ref{fig:VariousCrease}(d-e) reports the curvature distribution and the corresponding crease profiles by scaling the size of the creases in Figure \ref{fig:VariousCrease}(a) up to 150 times. In this case, significant differences among different creases are observed in the global shapes, and we would expect different mechanical behaviors of these structures subjected to identical loading conditions.

\begin{figure}[h!]
	\centering
	\includegraphics[width=0.9\textwidth]{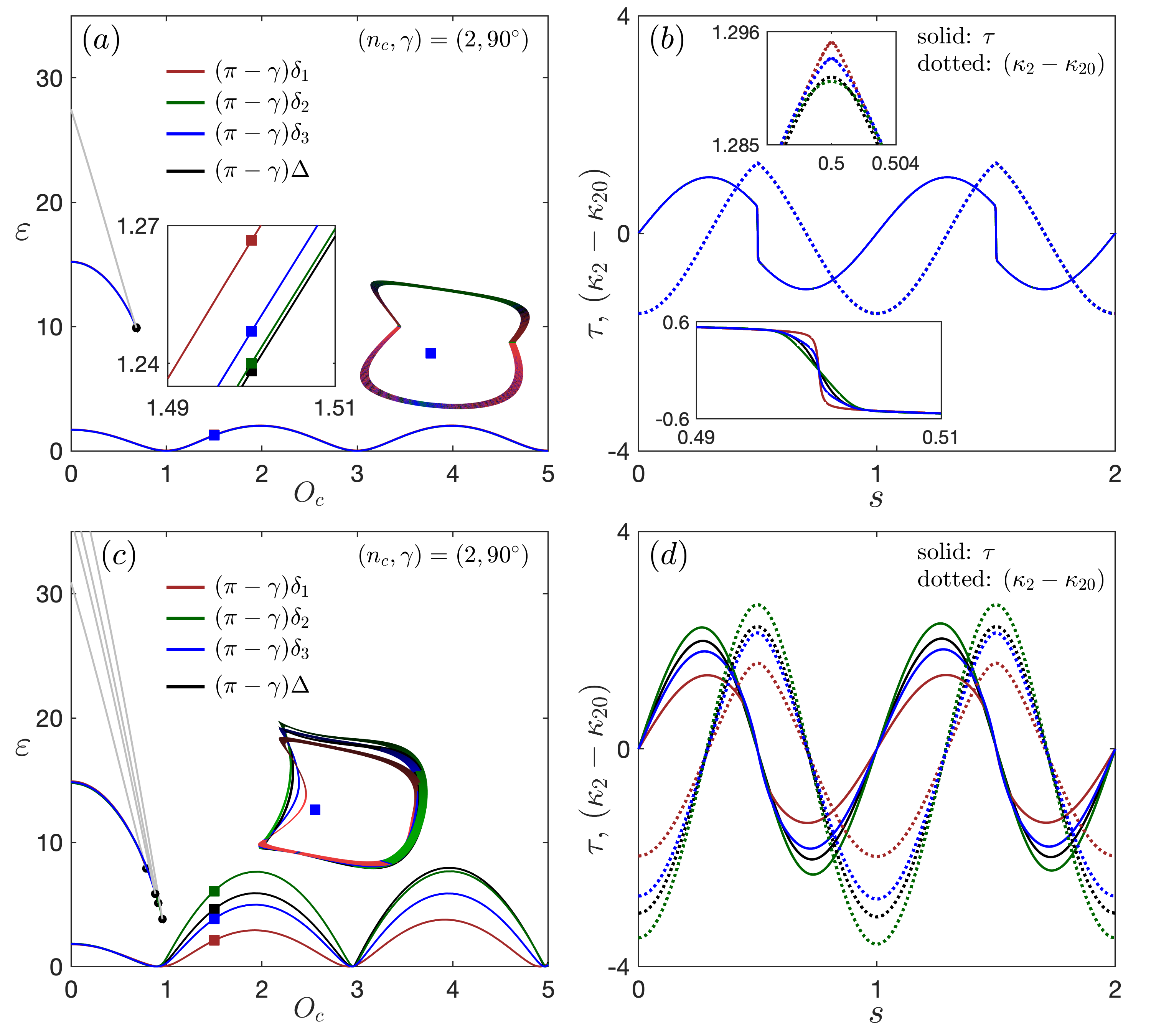}
	\caption{Numerical results of creased annular strips with $(n_c,\gamma)$ fixed to $(2,90^{\circ})$ and the creases described by different RDDFs in Equation \eqref{eq:DiracDelta}. (a) Solution curves of sharp creases with the same geometric parameters as in Figure \ref{fig:deltasAnnularN2}(a). 
		 (b) Strains of several equilibria corresponding to the squares in (a). (c) Solution curves of blunt creases with the same geometric parameters as in Figure \ref{fig:deltasAnnularN2}(d). 
		 (b) Strains of several equilibria corresponding to the squares in (c).} \label{fig:deltasAnnularN2}
\end{figure}

In the following, we use an example to demonstrate that as long as creases are sharp, any RDDF can be implemented to solve the nonlinear mechanics of creased structures.
Figure \ref{fig:deltasAnnularN2} displays numerical results of creased annuli with $(n_c,\gamma)$ fixed to $(2,90^{\circ})$ and the crease described by different RDDFs in Equation \eqref{eq:DiracDelta}. Figure \ref{fig:deltasAnnularN2}(a-b) include numerical results of sharp creases whose geometric parameters are the same as those in Figure \ref{fig:VariousCrease}(a). With sharp creases, the difference in the elastic energy of different crease profiles is negligible and the deformed shapes almost coincide with each other, demonstrated through the four square markers and their corresponding renderings. This is also true for the strains $\tau$ and $(\kappa_{2}-\kappa_{20})$ of the solutions marked by squares, reported in Figure \ref{fig:deltasAnnularN2}(b). 
The insets in Figure \ref{fig:deltasAnnularN2}(b) shows that notable differences only exist around the crease region.
In contrast, Figures \ref{fig:deltasAnnularN2}(c-d) present numerical results of blunt creases sharing the same geometric parameters with those in Figure \ref{fig:VariousCrease}(d). With different crease profiles, significant differences are found in the elastic energy $\varepsilon$, deformed shapes (the renderings in Figure \ref{fig:deltasAnnularN2}(c)), and strains (Figure \ref{fig:deltasAnnularN2}(d); corresponding to the squares in Figure \ref{fig:deltasAnnularN2}(c)).
We conclude that any RDDF could be used to model sharp creases without causing notable differences in the nonlinear mechanics (large deformations, stability, etc.) of creased strips. The $\Delta$ function proposed in the current study could address different types of discontinuities because it contains both Heaviside and Dirac-delta features.

\pagebreak
\newpage

\section{Slender rods with material discontinuities and nonlinear material properties} \label{appse:2Dstep} 

This study mainly focuses on the description of the sharp geometry of creases and assume the creased regions contain the same elastic materials as the rest part. 
However, creases could be more complicated in terms of both geometry and material properties. For example, without annealing, creases could follow a complex relaxation mechanism and have a highly nonlinear behavior \cite{thiria2011relaxation}. In order to promote the formation of crease, the thickness and width of the creased regions can be reduced \cite{shi2017plasticity,yan2016controlled,andrade2019foldable}, which leads to nonuniform cross section.

\begin{figure}[h!] 
	\centering
	\includegraphics[width=0.9\textwidth]{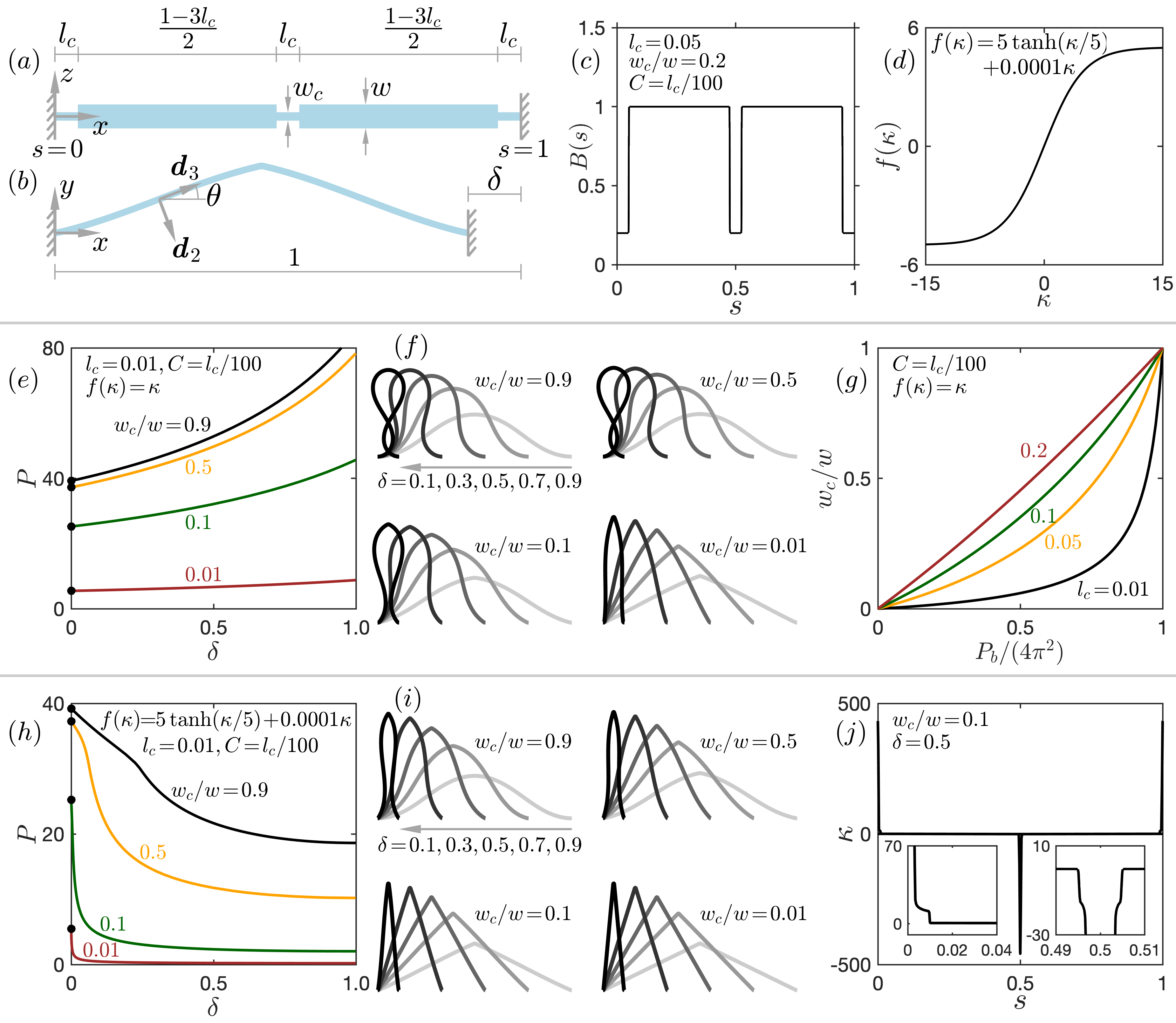}
	\caption{The formation of a crease. (a) The geometry of the ribbon in the rest configuration. The width is locally reduced at the mid and the two ends to promote folding. (b) The ribbon buckles out of plane with localized deformation at the width-reduction regions. (c) $B(s)$ accounts for the variation of width in (a) through a continuous function (\eqref{appeq:creasedelastica}). (d) $f(\kappa)$ follows a hyperbolic tangent plus a small linear term, mimicking an elastic-plastic constitutive law. The second row reports numerical results with $f(\kappa)=\kappa$. (e) Force-displacement curves with different $w_c/w$. The solid circles correspond to the critical load $P_b$ where the structure buckles out of plane. (f) Renderings of solutions in (e). (g) Loci of the critical load $P_b$ in the $w_c/w$ versus $P_b/(4 \pi^2)$ plane for different crease length $l_c$. The third row reports the numerical results with $f(\kappa)$ mimicking an elastic-plastic behavior. (h) Force-displacement curves with different $w_c/w$. (i) Renderings of solutions in (h). (j) Curvature distribution of a configuration with $w_c/w=0.1$ and $\delta=0.5$.}\label{fig:creaseformation}
\end{figure}

Here, we show an example using our framework to model a ribbon with the creased regions having a reduced width and the material following an approximate elastic-plastic behavior. We choose the example studied in \cite{shi2017plasticity}, where the authors show that both local reduction of width and plastic deformations can promote the formation of creases. Figure \ref{fig:creaseformation}a shows the geometry of the ribbon with both ends clamped and one end subject to a compressive displacement $\delta$ \cite{shi2017plasticity}. The total length of the rod is normalized to unity. $w_c$ and $w$ represent the reduced width at a crease and the width elsewhere, respectively. $l_c$ corresponds to the length of the creases, located in the regions $[0,l_c]$, $[l_c+ 0.5(1-3lc), 2l_c+ 0.5(1-3lc)]$, and $[1-l_c, 1]$. Upon compression, the system buckles out of plane, as shown in Figure \ref{fig:creaseformation}b. An orthonormal material frame $(\bm{d}_2,\bm{d}_3)$ is attached to the centerline of the rod; $\theta$ measures the angle between the unit tangent $\bm{d}_3$ and the horizontal direction.

For uniform rods made of linear elastic materials, the internal moment $M$ follows Hooke's law, i.e. $M(\kappa)=EI \kappa$. Here $E$ and $I$ corresponds to the Young's Modulus and the second moment of area, respectively. For general case, we assume 

\begin{equation}\label{appeq:Mks}
\begin{aligned}
M(\kappa,s) = B(s) f(\kappa) \,.
\end{aligned}	
\end{equation}

Through our continuous description of an arbitrary piecewise continuous function (see SI Appendix, section \ref{appse:NYCskyline}), $B(s)$ can account for the abrupt variation of cross sections, material thickness and material properties along the arc length and $f(\kappa)$ can characterize an arbitrary nonlinear dependence of bending moment on the curvature. 
In Figure \ref{fig:creaseformation}a, the abrupt change of the width leads to nonsmooth $B(s)$, which can be described as the following continuous function

\begin{equation}\label{appeq:creasedelastica}
\begin{aligned}
B(s) & = \frac{w_c}{w}  + (1.0-w_c/w) ( l_{e1}-l_{b1} ) \Delta_C ^{(l_{b1},l_{e1})} + (1.0-w_c/w) ( l_{e2}-l_{b2} ) \Delta_C ^{(l_{b2},l_{e2})} \\
\text{with} \,\, & l_{b1}=l_c, l_{e1}=l_c + \frac{(1-3l_c)}{2}, l_{b2}=2l_c + \frac{(1-3l_c)}{2}, l_{e2}=1-l _c \,.
\end{aligned}	
\end{equation}


We normalize $B(s)$ at a full-width cross section to unity. Figure \ref{fig:creaseformation}c shows an example of \eqref{appeq:creasedelastica}. When $C \to 0$, $B(s)$ will approach the ideal case. Here, we set $C=l_c/100$, which corresponds to a sharp jump of the cross section. Figure \ref{fig:creaseformation}d reports the profile of a hyperbolic tangent function plus a small linear term, mimicking an elastic plastic constitutive law.

We adopt the \emph{elastica} theory to solve this problem. The governing equations correspond to the planar version of \eqref{appeq:normF&M} and can be written as 

\begin{equation}\label{appeq:2Drodequiscalar}
\begin{aligned}
N_2' =N_3 \kappa \,, N_3' =-N_2 \kappa \,, \\
\kappa'  = (N_2-B'f)/(Bf_{\kappa}) \,, \\
\theta '=\kappa \,, x '= \cos \theta \,, y' = \sin \theta \,, s' = 1 \,, \\
\end{aligned}
\end{equation}

where a prime represents a partial derivative with respect to arc length $s$ and $f_{\kappa}=df/d \kappa$. $N_2$ and $N_3$ corresponds to the internal forces resolved on the material frame $(\bm{d}_2, \bm{d}_3)$. The differential equation for $\kappa$ is obtained by substituting $M=Bf$ into the moment balance equation $M'=N_2$. In addition, we introduce one more trivial equation $s' \!=\! 1$ because the arc length explicitly enters the equation through $M'$. The boundary conditions can be summarized as 

\begin{equation}\label{appeq:2DrodBCs}
\begin{aligned}
x(0) =0 ,  y(0) =0 , \theta(0)  = 0 , x(1) = 1- \delta, y(1) =0 , \theta(1)=0 , s(0)=0 \,. \\
\end{aligned}
\end{equation}

We conduct numerical continuation to solve the two point boundary value problem \eqref{appeq:creasedelastica}-\eqref{appeq:2Drodequiscalar}. The second row in Figure \ref{fig:creaseformation} presents the numerical results with $f(\kappa)=\kappa$. Figure \ref{fig:creaseformation}e reports the force displacement curves with different $w_c/w$ and $l_c$ fixed to 0.01. The solid circles correspond to the critical load $P_b$ where the structure buckles out of plane. With $w_c/w=1$, we obtain a buckling load 39.4784, which matches exactly with the Euler buckling load $4 \pi ^2$. Decreasing $w_c/w$ reduces the buckling load.

Figure \ref{fig:creaseformation}f shows a series of renderings from the solutions in Figure \ref{fig:creaseformation}e. Decreasing the crease width $w_c/w$ leads to more pronounced localized deformation at the creases and thus makes the facets flatter. However, even with a very small crease width $w_c/w=0.01$, bending of the facets are still noticeable. Figure \ref{fig:creaseformation}g shows the loci of the buckling point $P_b$ (normalized by Euler buckling load) in the $w_c/w$ versus $P_b/(4 \pi^2)$ plane for different crease length $l_c$. These loci curves are obtained by conducting two parameter continuation. Generally speaking, to achieve the same buckling load, a longer crease needs a larger width.

The third row reports numerical results with $f(\kappa)$ following a nonlinear elastic behavior, which mimics a material that approaches a plastic plateau stress after significant bending (see Figure \ref{fig:creaseformation}d). We wish to emphasize that here we are not attempting to model any unloading or recovery of the crease and only consider the forward formation of the crease, so that the distinction between elastic and plastic response is not an issue. 
Figure \ref{fig:creaseformation}h reports the force displacement curves with different $w_c/w$. Generic softening behaviors are observed, i.e. the load decreases after the buckling point. 
Once the plateau moment is reached at the crease, bending deformation quickly localizes there.
Figure \ref{fig:creaseformation}i shows a series of renderings from the solutions in Figure \ref{fig:creaseformation}h. Compared with Figure \ref{fig:creaseformation}f, the results in Figure \ref{fig:creaseformation}i clearly show that the introduction of a plateau bending moment can promote the formation of the crease, which behaves like a plastic hinge that can sustain large rotations with an almost constant bending moment. Notice that without much reduction of the crease width (see the case with $w_c/w=0.9$ and 0.5 in Figure \ref{fig:creaseformation}i), only introducing an elastic-plastic-like material law cannot guarantee flat facets.

Figure \ref{fig:creaseformation}j reports the curvature distribution of a configuration with $w_c/w=0.1$ and $\delta=0.5$. It is clearly that curvatures are highly localized at the creases. In addition, the two insets show vertical jump of curvature at the interface connecting the crease and the rest part of the ribbon (e.g. at $s=0.01$, $s=0.495$, and $s=0.505$), which is due to the fact that bending moment is continuous, but the bending stiffness jumps at the interface.

\section{Continuous description of 2D surfaces with geometric discontinuities} \label{2Dstep} 
	
\begin{figure}[h!]
	\centering
	\includegraphics[width=0.8\textwidth]{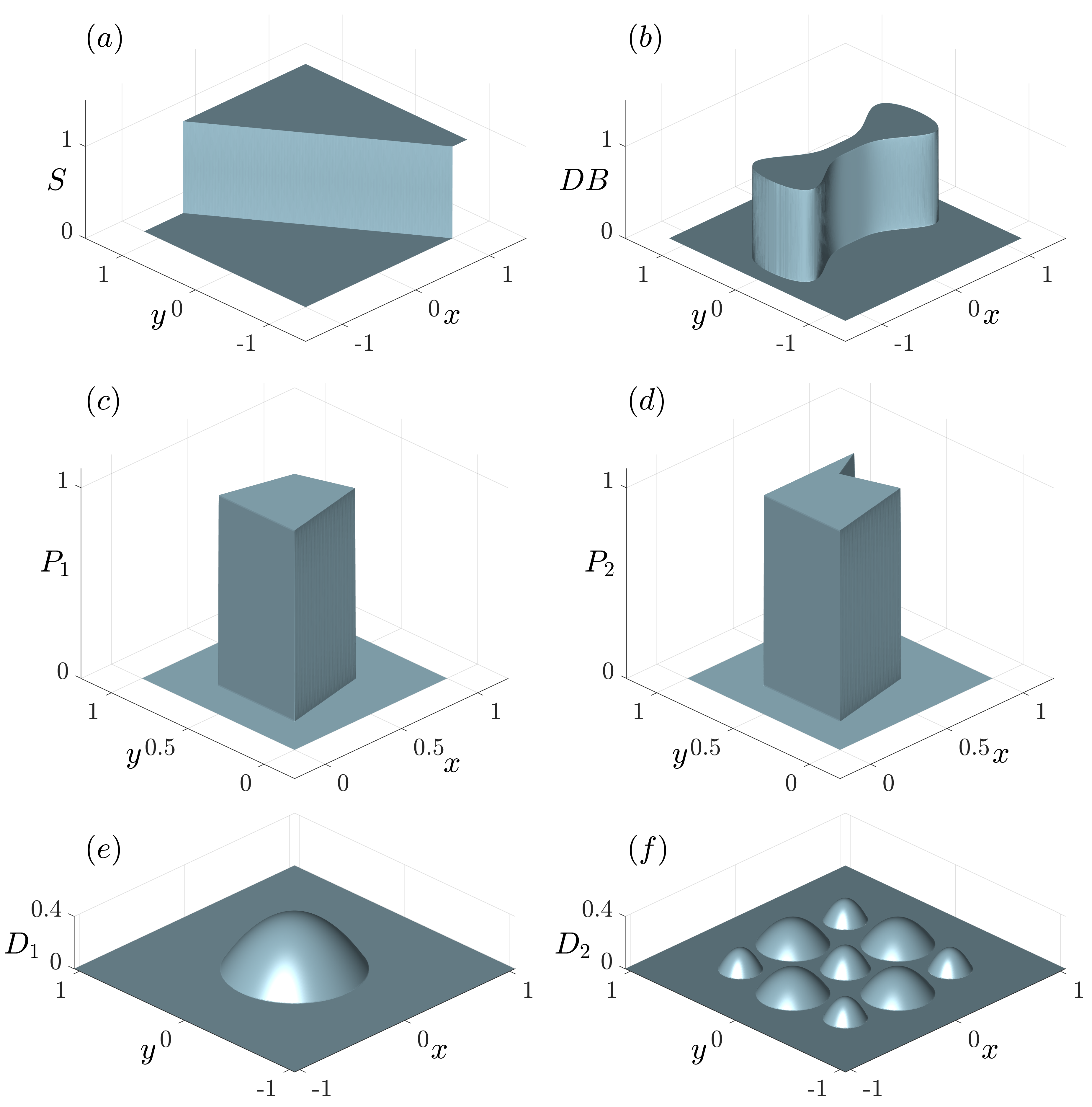}
	\caption{Different geometries based on the function in Equation \eqref{eq:Hxy} with $C$ fixed to 0.0002. (a) A half-plane based step $S$. (b) A dumbbell based step  $DB$. (c) A convex-polygon based step $P_1$. (d) A concave-polygon based step $P_2$. (e) A single dimple described by Equation \ref{appeq:D1}. (f) Dimples of different geometries in a thin sheet, described by Equation \ref{appeq:D2}. }\label{appfig:2Dstep}
\end{figure}

The creased annular strips studied in this work contain 1D discontinuities along the arc length of the strip. Here, we show that a hyperbolic tangent could be used to describe 2D surfaces with discontinuities. 2D steps can be described through the following equation	
	
\begin{equation}
	H(r(x,y))=\frac{1}{2} \left[  \tanh\left(\frac{r(x,y) }{C}\right) + 1 \right] \,,
	\label{eq:Hxy}
	\end{equation}

where $r$ is a real valued function of $x$ and $y$ and $C$ controls the sharpness of the step. When $C \to 0$, $H(x,y)=1$ for $r(x,y)>0$ and $H(x,y)=0$ elsewhere. The domain $r(x,y)>0$ will be referred to as the base of the 2D step and the curve $r(x,y)=0$ represents its boundary. Figures \ref{appfig:2Dstep}a and \ref{appfig:2Dstep}b displays a half-plane based step $S$ with $r=3x+2y-1$ and a dumbbell based step $DB$ with $r=x^4-x^6-y^2+0.01$, respectively. 
The product of several half-plane based step functions $H(r_i)$ can be used to describe a step with a convex polygonal base

\begin{equation}
P(x,y)=\prod_{i=1}^{n_s} H(r_i(x,y)) \,,
\label{eq:Pxy}
\end{equation}

where $n_s$ represents the number of sides of the polygonal base. The $i$-th side of the polygon base is part of the line $r_i(x,y)=0$. For a point inside the polygon $(x_p,y_p)$, we always have $r_i(x_p,y_p)>0$. Figure \ref{appfig:2Dstep}c displays a step with a convex polygonal base, described by the following function

\begin{equation}\label{eq:P1} 
\begin{aligned} 
P_1=&\left[ 0.5(\tanh( (0.5x-0.1)/C)+1) \right]
\left[ 0.5(\tanh( (-0.1x-0.4y+0.3)/C)+1) \right] \\
&\left[ 0.5(\tanh( (-0.3x-0.1y+0.24)/C)+1) \right]
\left[ 0.5(\tanh( (-0.1x+0.5y-0.08)/C)+1) \right] \, .
\end{aligned}
\end{equation}

Steps with a concave polygonal base can be obtained by adding those associated to the convex partitions of the concave base. Figure \ref{appfig:2Dstep}d presents an example combining the geometry in Figure \ref{appfig:2Dstep}(c) and an additional step with a convex polygonal base, described by the following function,

\begin{equation}\label{eq:P2} 
\begin{aligned} 
P_2=P_1 + & \left[ 0.5(\tanh( (-0.6y+0.42)/C)+1) \right] \left[ 0.5(\tanh( (-0.1x+0.2y-0.06)/C)+1) \right]\\
 & \left[ 0.5(\tanh( (0.1x+0.4y-0.3)/C)+1) \right] \, .
\end{aligned}
\end{equation}

Multiplying $H(x,y)$ by an arbitrary function allows us to generate a new function that equals to the original function within the base of $H$ and vanishes elsewhere. Figure \ref{appfig:2Dstep}e displays a single dimple on a thin sheet, described by

\begin{equation}\label{appeq:D1} 
\begin{aligned} 
D_1=\left[0.5( \tanh( (0.25-x^2-y^2)/C)+1) \right] \left[1.6  (0.25- x^2-y^2 )\right] \, .
\end{aligned}
\end{equation}

Combining several such functions, we are able to decorate multiple dimples of different geometries in a thin sheet. Figure \ref{appfig:2Dstep}f presents an example described by the following function

\begin{equation}\label{appeq:D2} 
\begin{aligned} 
D_2=& \left[0.5( \tanh( (0.17^2-x^2-y^2)/C)+1) \right] \left[6  (0.17^2- x^2-y^2 )\right] \\
+&\left[ 0.5( \tanh( (0.15^2-(x-0.5)^2-(y-0.5)^2)/C)+1) \right] \left[ 7(0.15^2- (x-0.5)^2-(y-0.5)^2 ) \right] \\
+&\left[ 0.5( \tanh( (0.15^2-(x+0.5)^2-(y+0.5)^2)/C)+1) \right] \left[ 7(0.15^2- (x+0.5)^2-(y+0.5)^2 ) \right] \\
+& \left[0.5( \tanh( (0.15^2-(x+0.5)^2-(y-0.5)^2)/C)+1)\right]  \left[7(0.15^2- (x+0.5)^2-(y-0.5)^2 ) \right]\\
+& \left[ 0.5 ( \tanh( (0.15^2-(x-0.5)^2-(y+0.5)^2)/C)+1)\right]  \left[7(0.15^2- (x-0.5)^2-(y+0.5)^2 ) \right]\\  
+& \left[ 0.5( \tanh( (0.25^2-(x-0.5)^2-y^2)/C)+1) \right] \left[3(0.25^2- (x-0.5)^2-y^2 ) \right] \\
+& \left[ 0.5 ( \tanh( (0.25^2-(x+0.5)^2-y^2)/C)+1) \right] \left[3(0.25^2- (x+0.5)^2-y^2 )\right] \\
+& \left[ 0.5 ( \tanh( (0.25^2-x^2-(y-0.5)^2)/C)+1) \right] \left[3 (0.25^2- x^2-(y-0.5)^2 ) \right]\\
+& \left[ 0.5 ( \tanh( (0.25^2-x^2-(y+0.5)^2)/C)+1) \right] \left[ 3(0.25^2- x^2-(y+0.5)^2 ) \right]\, .
\end{aligned}
\end{equation}

\begin{figure}[h!]
		\centering
		\includegraphics[width=0.6\textwidth]{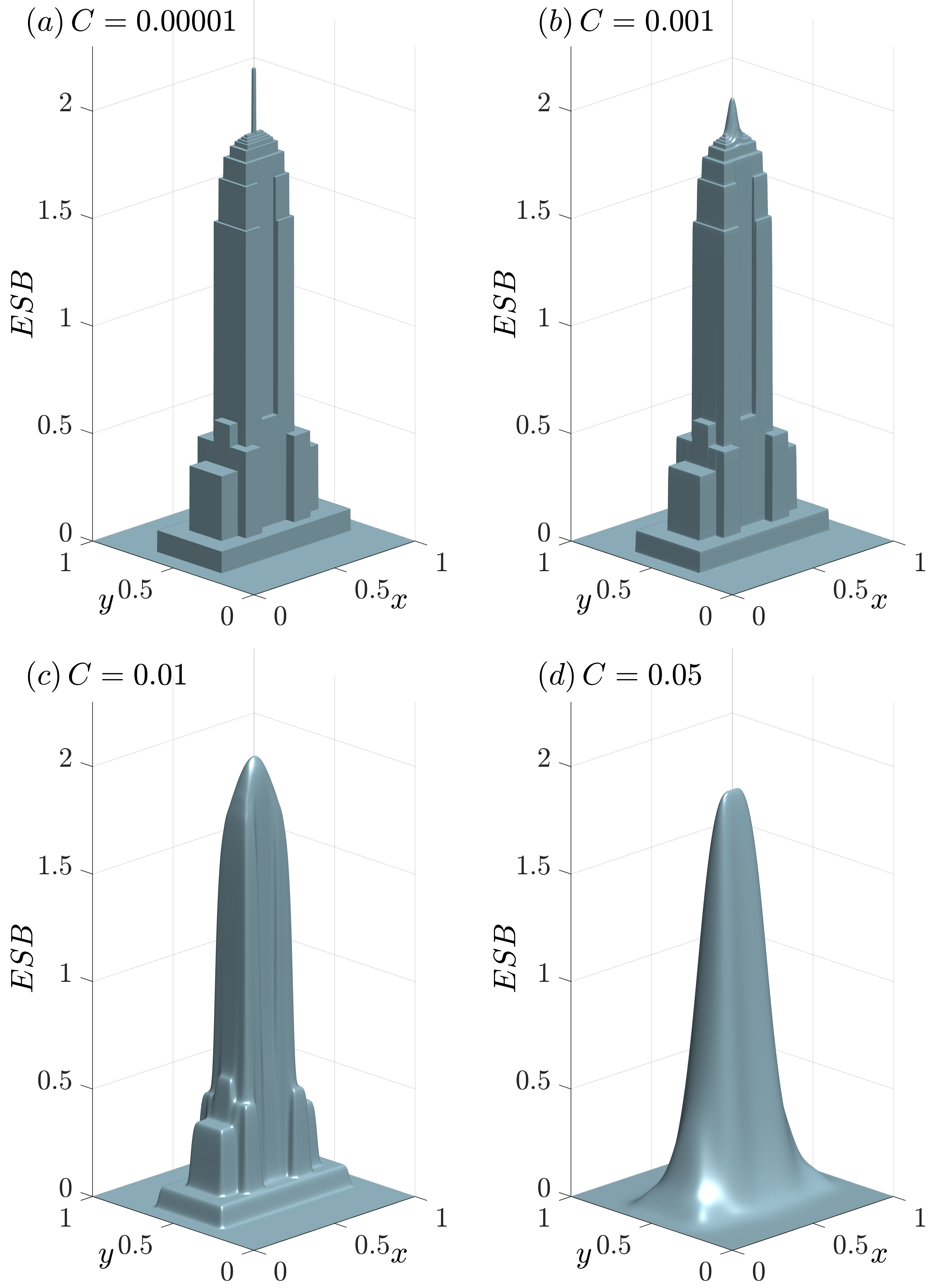}
		\caption{The external geometry of the Empire State Building characterized as a continuous surface by Equation \eqref{eq:ESB} with  (a) $C \!=\! 0.00001$, (b) $C \!=\! 0.001$, (c) $C \!=\! 0.01$, and (d) $C \!=\! 0.05$.}\label{fig:ESB}
	\end{figure}

The above continuous descriptions of 2D surfaces with discontinuities could be useful in numerical modeling of shape-morphing metasheets \cite{faber2020dome,liu2023snap}. In addition, $H$ can be used to approximate a complicated surface by partitioning it, describing each subdomain with a continuous function, and combining these functions together. Figure \ref{fig:ESB} displays the external surface of the Empire State Building, characterized by the following function

	\begin{equation} \label {eq:ESB}
	\begin{aligned} 
	ESB (x,y)= &\sum_{i=1}^{24} \frac{z_i}{2^4} 
	\left[\tanh\left(\frac{x - x_{i1}}{C}\right) + 1\right]
	\left[\tanh\left(\frac{x_{i2} - x}{C}\right) + 1\right]
	\left[\tanh\left(\frac{y-y_{i1}}{C}\right) + 1\right]
	\left[\tanh\left(\frac{y_{i2} - y}{C}\right) + 1\right] \\
	+&\frac{0.3}{2} \left  [  \tanh\left( \frac{0.01^2-(x - 0.5)^2 - (y - 0.5)^2}{C} \right) + 1\right   ]
	\end{aligned} 
	\end{equation}
	
	where the first 24 functions contain steps with rectangular bases $[x_{i1}, x_{i2}] \times [y_{i1},y_{i2}]$ and heights $z_i$. Their values are summarized in table \ref{tab:Empire}. The cylindrical antenna is modeled by the last term, corresponding to a circle based step. The plot in Figure \ref{fig:ESB}a is obtained with $C \!=\! 0.00001$, which plays an important role in determining the sharpness of the geometric discontinuity. Figures \ref{fig:ESB}(b-d) report this effect by rendering Equation \eqref{eq:ESB} with larger values of $C$.

\begin{table}[]
	\centering
	\begin{tabular}{|r|lllll||r|lllll|}\hline
		 & $x_{i1}$ & $x_{i2}$ & $y_{i1}$ & $y_{i2}$ & $z_i$ & $i$ & $x_{i1}$ & $x_{i2}$ & $y_{i1}$ & $y_{i2}$ & $z_i$\\ \hline
		$1$ & $0.1  $ & $0.9  $ & $0.3  $ & $0.7  $ & $0.1 $ & $13$ & $0.35 $ & $0.45 $ & $0.4  $ & $0.6  $ & $0.9 $ \\
		$2$ & $0.2  $ & $0.3  $ & $0.4  $ & $0.6  $ & $0.3 $ & $14$ & $0.55 $ & $0.65 $ & $0.4  $ & $0.6  $ & $0.9 $ \\
		$3$ & $0.7  $ & $0.8  $ & $0.4  $ & $0.6  $ & $0.3 $ & $15$ & $0.45 $ & $0.55 $ & $0.43 $ & $0.57 $ & $1.2 $ \\
		$4$ & $0.3  $ & $0.35 $ & $0.35 $ & $0.65 $ & $0.4 $ & $16$ & $0.365$ & $0.45 $ & $0.415$ & $0.585$ & $0.2 $ \\
		$5$ & $0.65 $ & $0.7  $ & $0.35 $ & $0.65 $ & $0.4 $ & $17$ & $0.55 $ & $0.635$ & $0.415$ & $0.585$ & $0.2 $ \\
		$6$ & $0.35 $ & $0.4  $ & $0.35 $ & $0.4  $ & $0.4 $ & $18$ & $0.38 $ & $0.45 $ & $0.43 $ & $0.57 $ & $0.1 $ \\
		$7$ & $0.35 $ & $0.4  $ & $0.6  $ & $0.65 $ & $0.4 $ & $19$ & $0.55 $ & $0.62 $ & $0.43 $ & $0.57 $ & $0.1 $ \\
		$8$ & $0.6  $ & $0.65 $ & $0.35 $ & $0.4  $ & $0.4 $ & $20$ & $0.4  $ & $0.6  $ & $0.45 $ & $0.55 $ & $0.05$ \\
		$9$ & $0.6  $ & $0.65 $ & $0.6  $ & $0.65 $ & $0.4 $ & $21$ & $0.42 $ & $0.58 $ & $0.46 $ & $0.54 $ & $0.02$ \\
		$10$ & $0.3  $ & $0.35 $ & $0.45 $ & $0.55 $ & $0.1 $ & $22$ & $0.43 $ & $0.57 $ & $0.47 $ & $0.53 $ & $0.01$ \\
		$11$ & $0.65 $ & $0.7  $ & $0.45 $ & $0.55 $ & $0.1 $ & $23$ & $0.44 $ & $0.56 $ & $0.48 $ & $0.52 $ & $0.01$ \\
		$12$ & $0.35 $ & $0.65 $ & $0.4  $ & $0.6  $ & $0.5 $ & $24$ & $0.45 $ & $0.55 $ & $0.49 $ & $0.51 $ & $0.01$ \\
		\hline
	\end{tabular}
	\caption{Parameters of the rectangle based step functions (see Equation \eqref{eq:ESB}) used to describe the Empire State Building in Figure \ref{fig:ESB}.}
	\label{tab:Empire}
\end{table}

\end{document}